\documentclass[english]{article}
\usepackage[T1]{fontenc}
\usepackage[utf8]{inputenc}
\usepackage{geometry}
\usepackage{color}
\usepackage{babel}
\usepackage{amsmath}
\usepackage{amsthm}
\usepackage{graphicx}
\usepackage{setspace}
\usepackage[unicode=true,pdfusetitle,
 bookmarks=true,bookmarksnumbered=false,bookmarksopen=false,
 breaklinks=false,pdfborder={0 0 1},backref=false,colorlinks=false]
 {hyperref}

\makeatletter
\usepackage{newunicodechar} 
\DeclareUnicodeCharacter{FFFD}{?????}

\makeatother

\begin{document}
\begin{doublespace}
\begin{center}
\textbf{\Large{}Behavioral Bias Benefits: Beating Benchmarks By Bundling
Bouncy Baskets}{\Large\par}
\par\end{center}

\begin{center}
\textbf{Ravi Kashyap (ravi.kashyap@stern.nyu.edu)}\footnote{\begin{doublespace}
Dr. Yong Wang, Dr. Isabel Yan, Dr. Vikas Kakkar, Dr. Fred Kwan, Dr.
William Case, Dr. Srikant Marakani, Dr. Qiang Zhang, Dr. Costel Andonie,
Dr. Jeff Hong, Dr. Guangwu Liu, Dr. Humphrey Tung and Dr. Xu Han at
the City University of Hong Kong; the editorial board, anonymous reviewers
and numerous seminar participants provided many suggestions to improve
this paper. The views and opinions expressed in this article, along
with any mistakes, are mine alone and do not necessarily reflect the
official policy or position of either of my affiliations or any other
agency.
\end{doublespace}
}
\par\end{center}

\begin{center}
\textbf{Estonian Business School, Tallinn, Estonia / Formation Fi,
Hong Kong / City University of Hong Kong, Hong Kong}
\par\end{center}

\begin{center}
\begin{center}
\today
\par\end{center}
\par\end{center}

\begin{center}
\par\end{center}

\begin{center}
and Effects of Psychological, Emotional, Social, and Cognitive Factors
on Decision Making ; G41 Role and Effects of Psychological, Emotional,
Social, and Cognitive Factors on Decision Making in Financial Markets
\par\end{center}

\begin{center}
\textbf{\textcolor{blue}{\href{https://doi.org/10.1111/acfi.12826}{Edited Version: Kashyap, R. (2021).  Behavioral Bias Benefits: Beating Benchmarks By Bundling Bouncy Baskets.  Accounting \& Finance,  XX(X),  XX-XX. }}}
\par\end{center}

\begin{center}
\tableofcontents{}
\par\end{center}
\end{doublespace}
\begin{doublespace}

\section{Abstract}
\end{doublespace}

\begin{doublespace}
We consider in detail an investment strategy, titled ``The Bounce
Basket'', designed for someone to express a bullish view on the market
by allowing them to take long positions on securities that would benefit
the most from a rally in the markets. We demonstrate the use of quantitative
metrics and large amounts of historical data towards decision making
goals. This investment concept combines macroeconomic views with characteristics
of individual securities to beat the market returns. The central idea
of this theme is to identity securities from a regional perspective
that are heavily shorted and yet are fundamentally sound with at least
a minimum buy rating from a consensus of stock analysts covering the
securities. We discuss the components of creating such a strategy
including the mechanics of constructing the portfolio. Using simulations,
in which securities lending data is modeled as geometric brownian
motions, we provide a few flavors of creating a ranking of securities
to identity the ones that are heavily shorted.

An investment strategy of this kind will be ideal in market scenarios
when a downturn happens due to unexpected extreme events and the markets
are anticipated to bounce back thereafter. This situation is especially
applicable to incidents being observed, and relevant proceedings,
during the Coronavirus pandemic in 2020-2021. This strategy is one
particular way to overcome a potential behavioral bias related to
investing, which we term the ``rebound effect''.\newpage{}
\end{doublespace}
\begin{doublespace}

\section{\label{sec:Trading-Strategy-Bounce}The Bounce Basket To Mitigate
The Market Rebound Effect}
\end{doublespace}

\begin{doublespace}
We consider in detail an investment strategy, titled ``The Bounce
Basket''. This strategy acts as a complete numerical illustration
of the general principles behind trading strategies discussed in Kashyap
(2019). We come up with a potential behavioral bias with respect to
the financial markets, which we term the ``rebound effect''. Our
investment strategy becomes one probable way to overcome such a bias.
We demonstrate the use of quantitative metrics and large amounts of
historical data towards decision making goals.

Our trading idea is designed for someone to express a bullish view
on the market by allowing them to take long positions on securities
that would benefit the most from a rally in the markets. This investment
concept combines macroeconomic views with characteristics of individual
securities to beat the market returns. The central idea of this theme
is to identity securities from a regional perspective that are heavily
shorted and yet are fundamentally sound (Summers 1986; Dechow, Hutton,
Meulbroek \& Sloan 2001; Bakshi \& Chen 2005) with at least a minimum
buy rating from a consensus of stock analysts covering the securities
(Barber, Lehavy, McNichols \& Trueman 2001; Barber, Lehavy, McNichols
\& Trueman 2003; Brown, Wei \& Wermers 2013).

The most heavily shorted securities are obtained from a top shorts
ranking model (Kashyap 2017b). To outline the intuition behind such
a ranking, suppose we wanted to rank the most livable cities in the
world (or let us say, the best soccer players in history). We would
identify characteristics that might be useful to make such an assessment.
We would weight the individual characteristics and then collect data
for the entities that are being ranked. We could then calculate an
overall score, which would be an aggregation based on the collected
data and the corresponding weights. The overall score would then yield
a suitable ranking, since having one number for each object we are
interested in facilitates a ranking. 

Once a ranking is formed, we could use some of the factors as the
criteria for exclusion from the finalized standings. These factors
act as filters to remove securities from the overall universe of stocks
in a region (say Asia) based on specific levels of the parameters.
Again, to outline the intuition here: a city could rank very well
on most factors, but could be below acceptable standards in one factor,
(say air quality), requiring that it be removed from the final ranking.
Hence, certain securities would be excluded from the final basket
if they fall below certain minimum acceptable thresholds. 

We relate our methodology to three different streams of literature
in Section (\ref{sec:Related-Literature-on}). Section (\ref{subsec:The-Market-Rebound})
has a discussion of one potential behavioral bias related to the financial
markets (but possibly lurking in many other areas of life), which
we term the ``rebound effect''. Section (\ref{sec:Securities-Lending-Factors})
provides a detailed discussion of the variables (or factors), specific
to securities lending that would be helpful to come up with a ranking
of the most heavily shorted securities, including one possible way
to weight the factors and combine them. Section (\ref{sec:Sharpening-the-Sharpe})
provides an alternative aggregation scheme that does not have subjective
weights as discussed in Section (\ref{sec:Securities-Lending-Factors}). 

Section (\ref{sec:Data-Generation-via}) shows how we generate sample
data using simulations in which securities lending information is
generated via geometric brownian motions (GBMs). The actual data for
back testing such a strategy would be available from the securities
lending desk of a broker dealer. We have taken great care to ensure
that the sample time series is similar to actual historical data in
terms of some of its statistical properties such as starting value,
drift and volatility (Equations: \ref{eq:6}; \ref{eq:7}).

Section (\ref{sec:Simple-Sample-Ranking}) illustrates numerical scores
for three flavors of the alternative aggregation scheme discussed
in Section (\ref{sec:Sharpening-the-Sharpe}) and shows how we can
obtain a top shorts ranking. Section (\ref{sec:Portfolio-Construction})
provides details on how to construct a portfolio based on the ranking
method in Section (\ref{sec:Sharpening-the-Sharpe}). Section (\ref{sec:Taming-the-Volatility})
has suggestions for improvement that address one of the main drawbacks
of using volatility as a decision making tool since it fails to fully
capture the effect of changes in the direction in the time series
of any variable. The figures for Sections (\ref{sec:Data-Generation-via};
\ref{sec:Simple-Sample-Ranking}; \ref{sec:Taming-the-Volatility})
are arranged in an Appendix (Section \ref{sec:Appendix-of-Figures})
according to the sections where they are referenced. These additional
figures should aid with better understanding of the concepts and the
related results discussed in the corresponding sections.

The investment time horizon for a trading strategy, such as the bounce
basket, will depend on the time horizon over which the markets are
expected to rebound and stay bullish. This would depend on the views
from global macro researchers that analyze countries, or regions,
and try to estimate whether the economic conditions are conducive
of growth combined with optimism about future prospects. A supplement
to the views of macro researchers would be an economic activity indicator
such as the Purchasing Manager's Index (Koenig 2002; Pelaez 2003;
Afshar, Arabian \& Zomorrodian 2007; Tsuchiya 2012)\footnote{Purchasing Managers' Indexes (PMI) are economic indicators derived
from monthly surveys of private sector companies. \href{https://en.wikipedia.org/wiki/Purchasing_Managers\%27_Index}{Purchasing Managers' Index, Wikipedia Link}}. 

Clearly, such views are bound to be erroneous and have a significant
degree of error. Hence, we could be conservative in our trading strategy
and the time horizon of our trade can be a fraction of the time over
which there is consensus expectation for the markets to trend upwards.
As an alternative to our investment idea, we could go long an index
future or exchange-traded fund\footnote{An exchange-traded fund (ETF) is a type of investment fund and exchange-traded
product, i.e. they are traded on stock exchanges. ETFs are similar
in many ways to mutual funds, except that ETFs are bought and sold
throughout the day on stock exchanges while mutual funds are bought
and sold based on their price at day's end. \href{https://en.wikipedia.org/wiki/Exchange-traded_fund}{Exchange Traded Fund, Wikipedia Link}} or take a derivative position on an index in anticipation of a market
rebound, which would be an example of a passive strategy.

The bounce basket is an example of an active strategy. In this case,
since we are carefully selecting the constituents of the basket from
a broader index or market, we can expect to outperform the overall
performance of the market. An investment strategy of this kind will
be ideal in market scenarios when a downturn happens, due to unexpected
extreme events, and the markets are anticipated to bounce back thereafter.
This situation is especially applicable to incidents being observed,
and relevant proceedings, during the Coronavirus pandemic in 2019,
2020 and 2021. Most studies focused on any crisis discuss how such
events adversely affect the financial markets. \textbf{\textit{We
wish to shed a ray of light in the current fight against COVID (and
also in other seemingly dire straits) by suggesting ways in which
investors can benefit and hope that this positivity will usher in
recoveries on all fronts (markets and otherwise).}}
\end{doublespace}
\begin{doublespace}

\subsection{\label{subsec:The-Market-Rebound}The Market Rebound Effect}
\end{doublespace}

\begin{doublespace}
A “rebound relationship” is commonly understood as a relationship
that is initiated shortly after a romantic breakup - before the feelings
about the former relationship have been resolved (Brumbaugh \& Fraley
2015). Research into the rebound effect, as it pertains to our love
lives, suggest that there might be some benefits to rebound relationships.
Spielmann, Macdonald \& Wilson (2009) demonstrate that focusing on
someone new may help anxiously attached individuals overcome attachment
to an ex-romantic partner, suggesting one possible motive behind so-called
rebound relationships. 

Barber \& Cooper (2014) find that, consistent with popular beliefs
about rebound and revenge sex, sexual episodes to cope with distress
and to get over or get back at the ex-partner were elevated immediately
following the breakup. Self-help books often advise readers to avoid
rushing into new relationships after a break-up. Wolfinger (2007)
tests the effects of rebound time, measured as time elapsed between
marital dissolution and the formation of a new union, on remarriage
duration and finds no evidence of a rebound effect. 

While the jury (in terms of undeniable research evidence) is still
out on the possibilities of rebound in our love lives, clearly, this
line of inquiry could be an interesting and fruitful pursuit in other
areas of life as well. We note that the rebound effect has another
meaning related to the consumption of energy services following an
improvement in the technical efficiency of delivering those services
(Berkhout, Muskens \& Velthuijsen 2000; Sorrell \& Dimitropoulos 2008;
Gillingham, et al., 2013). Though in our present context, we are concerned
with what people do once they experience a market crash and how they
react thereafter\footnote{Clearly, there are multiple ways of experiencing a rebound effect
with respect to financial investments. As an alternative to losses
caused due to a market crash, someone's portfolio might be exposed
only to a minor sub-section of the market and the value of their assets
might plummet significantly due to the nature of their specific holdings.
Many interesting questions and avenues of research arise here, in
terms of how people respond to casualties caused by different types
of investments, and how their later investment behavior is transformed
due to any mishaps experienced earlier.}.

With regards to the financial markets and investment decision making:
studying the timing, magnitude and types of investments people make
after facing losses - either due to bad selection (idiosyncratic risk)
or due to bad market events (systemic risk) - need to be pursued further\footnote{For detailed discussions on diversifiable and non-diversifiable risks,
see: Ross, et al., 2009}. If a rebound effect exists, with respect to investing and subsequent
to a market crash, people will load up on the entire market (or at-least
big portions of it) in the hopes of the market moving up. Hence, a
more careful selection of securities is bound to yield greater dividends.
Indiscriminate or haphazard investing on the back of a market crash,
which might have caused personal portfolios losses, would perhaps
not be advisable. A more circumspect approach can benefit from the
behavioral biases that are entrenched among the wider population and
are not easy to remedy (Section \ref{subsec:Biases-and-Investor}).
Even if the rebound effect does not exist, or is insignificant, our
strategy can still be used as a prudent means of stock selection that
can provide outsized returns. Our approach is certainly expedient
for someone who did not experience a market crash, but is eager to
make the most of an anticipated recovery in the markets.
\end{doublespace}
\begin{doublespace}

\section{\label{sec:Related-Literature-on}Related Literature on Crisis, Pandemics,
Biases, Behavior and Investments}
\end{doublespace}

\begin{doublespace}
We relate our paper to three different strands of literature: 1) Economic
/ financial crisis, pandemics and stock market reactions; 2) Behavioral
biases and decision making; and 3) Portfolio management techniques
and investment strategies.
\end{doublespace}
\begin{doublespace}

\subsection{\label{subsec:The-COVID-19-Pandemic}The COVID-19 Pandemic}
\end{doublespace}

\begin{doublespace}
Many excellent studies discuss the origins, transmission and other
characteristics of the COVID-19 infection and compare it to other
pandemics in recent recorded history (Guo, et al. 2020; Zhou, et al.
2020; Zhang, et al. 2020; Morens, et al. 2020; Zhang, Wu \& Zhang
2020; Shereen, et al. 2020; Petersen, et al.2020; Javelle \& Raoult
2020)\footnote{The COVID-19 pandemic, also known as the coronavirus pandemic, is
an ongoing pandemic of coronavirus disease 2019 (COVID-19) caused
by severe acute respiratory syndrome coronavirus 2 (SARS-CoV-2). \href{https://en.wikipedia.org/wiki/COVID-19_pandemic}{COVID-19 Pandemic, Wikipedia Link}}. No previous infectious disease outbreak, including the Spanish Flu,
has affected the stock market as forcefully as the COVID-19 pandemic.
COVID-19 has had significant impact on stock market volatilities across
the globe.

Mazur, Dang \& Vega (2020) investigate the US stock market performance
during the crash of March 2020 triggered by COVID-19 and find asymmetric
returns across sectors. They find that natural gas, food, healthcare,
and software stocks earn high positive returns, whereas equity values
in petroleum, real estate, entertainment, and hospitality sectors
fall dramatically. Moreover, loser stocks exhibit extreme asymmetric
volatility that correlates negatively with stock returns. 

Baek, Mohanty \& Glambosky (2020) focus on understanding regime changes
from lower to higher volatility using a Markov Switching Autoregressive
model. They explore the US stock market response to daily reporting
on COVID-19. Their results show that volatility is affected by specific
economic indicators and is sensitive to COVID-19 news. Both negative
and positive COVID-19 information is significant, though negative
news is more impactful, suggesting a negativity bias. Significant
increases in total and idiosyncratic risk are observed across all
industries, while changes in systematic risk vary across industry.
Though, as we discuss afterwards (Sections \ref{subsec:Markets-and-Crisis};
\ref{subsec:Biases-and-Investor}), there have been earlier inquiries
about overreaction related to unexpected negative news.

Baker, et al., (2020) suggest that government restrictions on commercial
activity and voluntary social distancing, operating with powerful
effects in a service-oriented economy, are the main reasons the U.S.
stock market reacted so much more forcefully to COVID-19 than to previous
pandemics in 1918--1919, 1957--1958, and 1968. Hanspal, Weber \&
Wohlfart (2020) survey a representative sample of US households to
study how exposure to the COVID-19 stock market crash affects expectations
and planned behavior. They provide correlational and experimental
evidence that beliefs about the duration of the stock market recovery
shape households’ expectations about their own wealth and their planned
investment decisions and labor market activity. Wealth shocks are
associated with upward adjustments of expectations about retirement
age, desired working hours, and household debt, but have only small
effects on expected spending. 

Phan \& Narayan (2020) observe 25 countries’ stock market data and
provide a preliminary discussion on how the most active financial
indicator -- namely, the stock price -- reacted in real-time to
different stages in COVID-19’s evolution. They argue that, as with
any unexpected news, markets over-react and as more information becomes
available and people understand the ramifications more broadly the
market corrects itself. Their hypothesis requires further empirical
\& more robust verification.
\end{doublespace}
\begin{doublespace}

\subsection{\label{subsec:Markets-and-Crisis}Financial Markets and Crisis}
\end{doublespace}

\begin{doublespace}
Looking beyond the COVID-19 pandemic, we consider the broader linkages
between financial / economic crisis and financial markets. Kindleberger
\& Aliber (2011) is an entertaining account of the history of crises
and speculative manias. Reinhart \& Rogoff (2009) prove wrong many
claims that the old rules of valuation no longer apply and that the
new situation bears little similarity to past disasters. Odekon (2015)
is a timely and authoritative exploration of three centuries of good
times and hard times in major economies throughout the world. Events
from Tulipmania during the 1630s to the US federal stimulus package
of 2009 are covered. 

Thomas \& Morgan-Witts (2014) is the story of an overheated stock
market and the Wall Street crash of 1929, that led to the Great Depression
of the 1930s. Ohanian (2009) develops a theory of labor market failure
for the Depression based on Hoover's industrial labor program that
provided industry with protection from unions in return for keeping
nominal wages fixed. Richardson \& Troost (2009) consider the varying
monetary interventions by different federal reserve banks during the
1930 banking crisis. Atlanta expedited lending to banks in need, while
St. Louis did not. In Atlanta, banks survived at higher rates, lending
continued at higher levels, commerce contracted less, and recovery
began earlier. These patterns indicate that central bank intervention
influenced bank health, credit availability, and business activity.

Granados \& Roux (2009) used historical life expectancy and mortality
data to examine associations of economic growth with population health
for the period 1920--1940. They find that population health did not
decline and indeed generally improved during the 4 years of the Great
Depression, 1930--1933, with mortality decreasing for almost all
ages, and life expectancy increasing by several years in males, females,
whites, and nonwhites. Eichengreen \& Irwin (2010) note that the Great
Depression was marked by a severe outbreak of protectionist trade
policies. But contrary to the presumption that all countries scrambled
to raise trade barriers, there was substantial cross-country variation
in the movement to protectionism.

A series of studies have been performed on the co-movement among the
stock markets since the so-called October crash of 1987, when the
stock markets in the world collapsed globally one after another. Roll
(1988) found that all major world markets declined substantially in
October 1987 (of 23 markets, 19 declined more than 20 per cent). This
is an exceptional occurrence, given the usual modest correlations
of returns across countries observed at that time. Schwert (1990)
analyzed the behavior of stock return volatility, using daily data
from 1885 through 1988, and found that stock volatility jumped dramatically
during and after the crash\footnote{October 19 was the largest percentage change in market value in over
29,000 days. }. But he confirms, using implied volatilities from call option prices
and estimates of volatility from futures contracts on stock indexes,
that volatility returned to lower, more normal levels more quickly
than past experience predicted. 

Lee \& Kim (1993) find that national stock markets became more interrelated
after the 1987 crash, and the strengthening co‐movements among national
stock markets continued for a longer period after the crash. Choudhry
(1996) studied volatility, risk premia and the persistence of volatility
in six emerging stock markets before and after the 1987 stock market
crash. Using monthly data, from Argentina, Greece, India, Mexico,
Thailand, and Zimbabwe between January of 1976 and August of 1994,
he noted that the changes were not uniform and they depended upon
the individual markets including the possibility that factors other
than the 1987 crash may also be responsible for the changes.

The Asian financial crisis, triggered by the collapse of the value
of Thai Baht in July 1997, spread to all the countries in the region.
Khan \& Park (2009) present empirical evidence of herding contagion
in the stock markets during the 1997 Asian financial crisis, above
and beyond macroeconomic fundamental driven co-movements\footnote{Khan \& Park (2009) define contagion as a significant increase in
cross-market linkages after an initial shock to one country or a group
of countries. Within this framework, test for contagion boils down
to verifying if the cross-market co-movements increase significantly
after a shock. The argument is: if correlations increase significantly
in the crisis period compared to the tranquil period, one may conclude
in favor of herding contagion. This happens because international
financial markets tend to move more closely together during a period
of turbulence. Due to its simplicity, this approach has become a relatively
standard tool in the literature on contagion.}. 

Jang \& Sul (2002) select seven stock markets: Thailand, Indonesia
and Korea as direct crisis countries and Japan, Hong Kong, Singapore
and Taiwan as neighboring countries. Then dividing the period of study
into three sub-periods of pre-crisis, crises and post-crisis period,
they try to answer such questions as whether there existed a common
trend in Asian stock markets before the crisis, whether the correlations
among the Asian stock markets are increased due to the crisis and
whether there are any changes in the causal relations between the
stock markets during the sample period. They use time series tests
such as the co-integration test and the Granger causality test (Granger
1969).

Nikkinen, Piljak \& Äijö (2012) investigate linkages between developed
European stock markets and emerging stock markets. They focus on three
countries in the Baltic region, namely Estonia, Latvia and Lithuania
with particular attention to the recent financial crisis of 2008--2009.
They demonstrate that the Baltic stock markets were apparently segmented
before the crisis and that they were highly integrated during the
crisis. This suggests that there are less diversification benefits
during crises when investors would need them the most.
\end{doublespace}
\begin{doublespace}

\subsection{\label{subsec:Analysis-of-Investment}Analysis of Investment Strategies}
\end{doublespace}

\begin{doublespace}
Clearly, there is a strong line of inquiry in the existing literature
related to better portfolio management techniques and the analysis
of investment strategies. (Kashyap 2019) has a detailed discussion
of some general principles that can be utilized towards goals of seeking
excess returns.

Browne (2000) considers a dynamic active portfolio management problem
where the objective is related to the tradeoff between the achievement
of performance goals and the risk of a shortfall. Specifically, the
objective relates the probability of achieving a given performance
objective to the time it takes to achieve the objective. Most studies
that analyze dynamic investment strategies have obtained explicit
results by restricting utility functions to a few specific forms.
The resultant dynamic strategies have exhibited a very limited range
of behavior, not surprisingly. In contrast, Cox \& Leland (2000) consider
any specific dynamic strategy and ask whether we can characterize
the results of following it through time? More precisely, they try
to determine whether it is self-financing, yields path-independent
returns, and is consistent with optimal behavior for some expected
utility maximizing investor. They provide necessary and sufficient
conditions for a dynamic strategy to satisfy each of these properties.
Ammann, Kessler \& Tobler (2006) use tracking error variance (TEV)
as a measure of activity and introduce two decompositions of TEV for
identifying different investment strategies

Ranaldo \& Haeberle (2008) argue that the commonly used market indices
imply forms of active investment management in disguise. Malkiel (2003)
presents the case for and the evidence in favor of passive investment
strategies and examines the major criticisms of the technique. Balvers,
Wu \& Gilliland (2000) find strong evidence of mean reversion in relative
stock index prices. They use additional cross-sectional power gained
from national stock index data of 18 countries during the period 1969
to 1996. Their findings, which are robust to alternative specifications
and data, imply a significantly positive speed of reversion with a
half-life of three to three and one-half years. They devise a parametric
contrarian strategy, which efficiently exploits the information on
mean reversion across countries directly from the parameter estimates
of their econometric model, that outperforms buy-and-hold and standard
contrarian strategies.\footnote{Balvers, Wu \& Gilliland (2000) use the term “contrarian strategy”
in its general sense, as signifying buying (selling) assets that have
performed poorly (well) in the past. The standard DeBondt \& Thaler
(1985) zero-net-investment strategy (short-selling assets that have
performed well and using the proceeds to buy assets that have performed
poorly) would then become just a particular example of a contrarian
strategy. The term “momentum strategy” correspondingly has the opposite
meaning.}
\end{doublespace}
\begin{doublespace}

\subsection{\label{subsec:Biases-and-Investor}Biases, Games and Investor Behavior}
\end{doublespace}

\begin{doublespace}
There are numerous studies that investigate the effects of behavior
and biases with respect to investing. Investor sentiment, defined
broadly, is a belief about future cash flows and investment risks
that is not justified by the facts at hand. Pompian (2011) is an excellent
guide for understanding how to use behavioral finance theory in investing.
Menkhoff \& Nikiforow (2009) provide evidence on the hypothesis that
many behavioral finance patterns are so deeply rooted in human behavior
that they are difficult to overcome by learning.

De Bondt \& Thaler (1985) perform a study of market efficiency that
investigates whether the tendency for most people to “overreact” to
unexpected and dramatic news events affects stock prices. They present
empirical evidence using monthly data for the period between January
1926 and December 1982, from the Center for Research in Security Prices
(CRSP) at the University of Chicago, which is consistent with the
overreaction hypothesis.

The question is no longer whether investor sentiment affects stock
prices, but how to measure investor sentiment and quantify its effects.
One approach is \textquotedbl bottom up'' - using biases in individual
investor psychology, such as overconfidence, representativeness, and
conservatism - to explain how individual investors under-react or
overreact to past returns or fundamentals. Baker \& Wurgler (2007)
develop a \textquotedbl top down\textquotedbl{} and macroeconomic
approach. They take the origin of investor sentiment as exogenous
and focus on its empirical effects. They show that it is quite possible
to measure investor sentiment and that waves of sentiment have clearly
discernible, important, and regular effects on individual firms and
on the stock market as a whole. This approach builds on the two broader
and more irrefutable assumptions of behavioral finance -{}- sentiment
and the limits to arbitrage -{}- to explain which stocks are likely
to be most affected by sentiment. They suggest in particular that
stocks that are difficult to arbitrage or to value are most affected
by sentiment.

Oechssler, Roider\& Schmitz (2009) provide an experimental test for
the hypothesis that the incidence of behavioral biases is related
to cognitive abilities. They find that individuals with low cognitive
reflection test (Frederick 2005) scores are significantly more likely
to be subject to the conjunction fallacy (Tversky \& Kahneman 1983)
and to conservatism with respect to probability updating. Loewenstein
(2000) argues that emotions are not only important but the determinations
of emotional factors and their impact on economic behavior are amenable
to formal modeling. van den Bergh \& Gowdy (2009) examine the role
of group dynamics and interactions in explaining economic behavior.

Kudryavtsev, Cohen \& Hon-Snir (2013) analyze the effects of five
well-documented behavioral biases\footnote{A more complete list of cognitive biases is given here: \href{https://en.wikipedia.org/wiki/List_of_cognitive_biases}{List of Cognitive Biases, Wikipedia Link}}
- namely, the disposition effect, herd behavior, availability heuristic,
gambler’s fallacy and hot hand fallacy - on the mechanisms of stock
market decision making and, in particular, the correlations between
the magnitudes of the biases in the cross-section of market investors.
Employing an extensive online survey, they demonstrate that, on average,
active capital market investors exhibit moderate degrees of behavioral
biases. By calculating the cross-sectional correlation coefficients
between the biases they find that all of them are positive and highly
significant for both professional and non-professional investors and
for all categories of investors, as classified by their experience
levels, genders, and ages.

Game theoretical techniques (Fudenberg \& Tirole 1991; Gibbons 1992)
can be used to understand the decision strategies of individuals.
While such an approach can be insightful to understand decision making
with limited participants, extending those results to markets is not
a straightforward affair. Payne, Laughhunn \& Crum (1980) present
model of the effects of a reference point on risky choice behavior.
Target returns and reference points represent variations on the concept
of an aspiration level, an old idea in theories of decision making.
They present additional evidence on the need to incorporate such a
concept in the analysis of risky choice behavior.

Charness, Cobo-Reyes \& Jiménez (2008) explore the effect of the possibility
of third-party intervention on behavior. A third-party's material
payoff is not affected by the decisions made by the other participants,
but this person may choose to punish a responder who has been overly
selfish. The concern over this possibility may serve to discipline
potentially selfish responders. Buchan, Croson \& Solnick (2008) pose
the question - does gender influence trust, the likelihood of being
trusted and the level of trustworthiness? They compare choices by
men and women in an investment game and use questionnaire data to
try to understand the motivations for the behavioral differences.
They find that men trust more than women, and women are more trustworthy
than men. Abbink, Irlenbusch \& Renner (2000) introduce a game (Berg,
Dickhaut \& McCabe 1995) that allows the study of both positively
and negatively reciprocal behaviour.
\end{doublespace}
\begin{doublespace}

\section{\label{sec:Securities-Lending-Factors}Securities Lending Factors}
\end{doublespace}

\begin{doublespace}
Below we discuss some of the factors that can be helpful towards forming
a top shorts ranking from a securities lending perspective and the
rationale behind the usage of these factors to create the ranking.
We also provide contextual information and numerical pointers that
could be used to implement the filtering mechanism. 
\end{doublespace}
\begin{enumerate}
\begin{doublespace}
\item \label{enu:Short-Interest-(SI)}Short Interest ($SI$) - The short
interest is the amount of shares shorted in a security by all market
participants and is clearly a direct indicator of investors intending
to express negative sentiment on that stock. This metric is best expressed
in USD so it is normalized across securities and also when dealing
with multi-market regions like Asia. Only securities with Short Interest
of more than a certain USD value (say 10 million) are considered for
inclusion into the basket. An adjustment is made to factor in the
relative size of the markets. This would mean that for two securities
with say short interest of 11 million USD, one from Japan and one
from Taiwan, the Taiwan security would get ranked higher since the
stock loan market for Japan is higher and the levels of short interest
are much higher for Japanese stocks (Asquith, Pathak \& Ritter 2005;
Kashyap 2016; Bris, Goetzmann \& Zhu 2007)\footnote{\begin{doublespace}
\noindent SI can also be expressed as a ratio, shown below, but using
the actual share amount makes for a more granular ranking and generally
traders want to know how many shares are shorted compared not just
to the float amount, but the supply they can obtain from other lenders.
Also the loan rates are higher when the short interest compared to
the float is higher, so another factor captures the effect of this
ratio. The SI ratio is also included in our model as a separate factor
(Point \ref{enu:Days-To-Cover}) referred to below as the Days to
Cover.

\noindent The short interest ratio (also called days-to-cover ratio)
represents the number of days it takes short sellers on average to
cover their positions, that is repurchase all of the borrowed shares.
It is calculated by dividing the number of shares sold short by the
average daily trading volume, generally over the last 30 trading days.
The ratio is used by both fundamental and technical traders to identify
trends. The days-to-cover ratio can also be calculated for an entire
exchange to determine the sentiment of the market as a whole. If an
exchange has a high days-to-cover ratio of around five or greater,
this can be taken as a bearish signal, and vice versa. \href{https://en.wikipedia.org/wiki/Short_interest_ratio}{Short Interest Ratio, Wikipedia Link}
\end{doublespace}
} ).
\item Loan Rates ($LR$) - The most important variable that indicates how
heavily shorted a security is, is the loan rate. Only securities with
annual loan rates greater than a threshold (say 1.5 percent) are considered.
An adjustment is done to factor in the maximum level of loan rates
in each market similar to Point (\ref{enu:Short-Interest-(SI)}) above.
A lending desk has at-least two flavors of these loan rates: a rate
at which the desk is able to source inventory and another alternate
rate which is slightly higher since it is the rate at which loans
are made to end investors. To avoid any confusion, we refer to these
as the loan rate and the alternate loan rate respectively.
\item \label{enu:Days-To-Cover}Days To Cover ($DTC$) - Days to cover (also
known as the Short Interest Ratio: Hong, Li, Ni, Scheinkman \& Yan
2015; Point \ref{enu:Short-Interest-(SI)}) is a measurement of a
company's issued shares that are currently shorted, expressed as the
number of days required to close out all of the short positions. It
is calculated by taking the number of shares that are currently shorted
and dividing that amount by the average daily volume for the shares
in question. For example, if a company has average daily volume of
1 million shares and 2 million shares are currently short sold, the
shares have a cover rate of 2 days (2 Million/1 Million). A higher
value for this metric indicates that if investors wish to cover their
shorts, it could take many days and during that time, their positions
could end up losing money significantly. Only securities requiring
more than a certain days to cover, (say 4) are considered.
\item Loan Balance Growth ($LBG$) - The loan balance is a notional amount
(measured in USD) and represents the total amount of loans being made
by a particular broker or short selling desk. Hence, growth in this
amount (expressed in percentage terms), captures the increase in the
short sentiment being seen by this participant. This information is
sometimes shared with clients who are interested in trading with the
desk. Only securities with loan balance growth of more than a certain
percentage, (say 25 percent) over the past few months are considered.
\item Inverse Loan Availability ($ILA$ or Loan Availability, $LA$) - The
availability is the amount of shares that could be possibly borrowed
towards putting on additional short positions. It is the amount available
for shorting but currently unused. The inverse of this number (normalized
by expressing in USD) is used as contribution factor towards a total
score, since lower this amount, greater the pressure on loan rates
and a higher indication that a particular security is heavily shorted.
Only securities with loan availability below a certain threshold are
considered (say 10 million USD) are considered.
\item Liquidity ($L$) - We consider the average daily trading volume to
ensure that once a basket is constructed, we have enough volume being
traded so that we can implement this strategy without liquidity problems.
Only securities that had a 20 day average trading volume ($ADV$)
of more than a certain figure (say 25 million USD) are considered.
\item Buy Rating $(BR$) - Only securities with a minimum buy rating from
a consensus of analysts covering the stocks are considered. The rationale
behind this is that, as markets rebound, the securities that are fundamentally
sounds are the ones that will receive positive inflows and trend up
higher. Securities that have problems with their operations are likely
to remain in the negative investor sentiment zone, or stay shorted.
\item High Beta ($HB$) - The securities with a relatively high beta in
comparison to the market are chosen so that they are geared to capitalize
the most from any upward movement in the overall market.  Only Securities
with Beta more than a certain value (say 1.2) are considered. 
\end{doublespace}
\end{enumerate}
\begin{doublespace}
The factors related to securities lending$\left(SI,LR,DTC,LBG,ILA\right)$
and the liquidity$\left(L\right)$ are technical in nature. That is,
they have a time series that changes quite often. Hence they have
to be further checked for stability and changes over time by looking
at the moving average and volatility of the particular factor over
the last few (two or three) months. Securities with a higher moving
average are ranked higher and securities with higher volatility are
ranked lower. 

The variables we have discussed above are easier to acquire and utilize
for investment firms. Most broker dealers are likely to act as good
sources of information for these variables. Some of the additional
factors we outline next are hard even for broker dealers to estimate
accurately. Close attention also needs to be paid to regulatory restrictions
that might differ by regions and how they might affect the variables
that are under consideration. The tradeoffs between the costs to acquire
additional information and the potential benefits they bring need
to be assessed carefully. 

Additional factors such as the total number of lenders, the number
of total shares available for securities loans, etc. can be included.
As the number of lenders or market participants, making loans or having
short positions on a particular security respectively, increases the
effect each of them will have on the overall movement in the variables
will decrease. Though with more players, there is likely to be more
uncertainty. Generally, the number of lenders is inversely related
to the loan rates, since having more sources of supply can mitigate
the effects of a few sources drying up. As the short interest gets
closer to the total shares available for loans, loan rates might skyrocket
and become highly volatile. 

The factor weights are estimated using a trial and error approach
(Ismail 2014; Kashyap 2021)\footnote{\begin{doublespace}
\noindent As Taleb explains, ``it is trial with small errors that
are important for progress''. The emphasis on small errors is especially
true in a portfolio management context, since a huge error could lead
to a blow up of the investment fund. (Ismail 2014) mentions the following
quote from Taleb, “Knowledge gives you a little bit of an edge, but
tinkering (trial and error) is the equivalent of 1,000 IQ points.
It is tinkering that allowed the industrial revolution''. \href{https://www.youtube.com/watch?v=MMBclvY_EMA}{Link for Nassim Taleb and Daniel Kahneman discussing Trial and Error / IQ Points, among other things, at the New York Public Library on Feb 5, 2013.}
\end{doublespace}
} so that the profitability of the securities chosen from a securities
lending perspective over a historical period are maximized. The rationale
for such a weighting scheme is that a security is the most shorted
security if it was creating significant profits for the trading desk.
One sample model for the overall scoring mechanism can be summarized
as shown in Equation (\ref{eq:1}). Here $n$ is the total number
of factors. $w_{i}$ is the weights of the individual factors, $f_{i}$.

\begin{align}
\text{Total Score} & =\sum_{i=1}^{n}w_{i}f_{i}=w_{si}SI+w_{lr}LR+w_{dtc}DTC+w_{lbg}LBG+w_{ila}ILA\label{eq:1}\\
 & +\text{ Other factors to capture moving average and volatility of individual variables}\nonumber \\
\text{Also, } & \sum_{i=1}^{n}w_{i}=1\nonumber 
\end{align}

\end{doublespace}
\begin{doublespace}

\section{\label{sec:Sharpening-the-Sharpe}Sharpening the Sharpe Ratio}
\end{doublespace}

\begin{doublespace}
A key issue with the above scoring mechanism (Equation: \ref{eq:1}
in Section \ref{sec:Securities-Lending-Factors}) to create a top
shorts ranking is the subjectivity of the weights and the necessity
of having to constantly tweak them. A significant amount of effort
is involved to ensure that the weights are closely aligned with the
drivers of the Profit and Loss (P\&L). As an alternate to the above
scoring mechanism, which is an aggregation of a number of separate
factors weighted subjectively, we could use a single numeric indicator
similar to the Sharpe ratio, SR (Sharpe 1966; 1994)\footnote{\begin{doublespace}
\noindent In finance, the Sharpe ratio (also known as the Sharpe index,
the Sharpe measure, and the reward-to-variability ratio) is a way
to examine the performance of an investment by adjusting for its risk.
The ratio measures the excess return (or risk premium) per unit of
deviation in an investment asset or a trading strategy, typically
referred to as risk, named after William F. Sharpe. \href{https://en.wikipedia.org/wiki/Sharpe_ratio}{Sharpe Ratio, Wikipedia Link}
\end{doublespace}
}. 

The SR combines the risk, ($\sigma_{r}$, the standard deviation of
returns), and return, ($E\left(r\right)$, the expected return or
an estimate based on historical returns), of each individual asset
or portfolio to give one number that contributes to a ranking. The
return figure used is the return of the asset in excess of a certain
threshold, which is usually the risk free rate, $r_{f}$, but it could
be the return on any other benchmark security ($SR=\frac{E\left(r\right)-r_{f}}{\sigma_{r}}$
). This ensures that assets providing returns below the threshold
have a negative score and are relegated towards the bottom of the
ranking. 

A similar score can be created based on the loan rates and the corresponding
standard deviation as shown in Equation (\ref{eq:2}). The key point
to note here is that we are considering the expectation of loan rates
and the corresponding standard deviation of the loan rates instead
of the price returns and the standard deviation of the returns, which
are used in the SR.

\begin{equation}
\text{Short Score One}=\frac{E\left(LR\right)-r_{f}}{\sigma_{LR}}\label{eq:2}
\end{equation}

If the loan rates for certain securities are below the risk free rate
(or another higher threshold), any loans made on those securities
are not contributing much to bottom-line. Positions on such securities
could be available internally, since other trading desks within the
firm could be long those stocks. To obtain those shares the firm would
need to borrow money at the risk free rate (or higher), but making
securities loans on those positions will only earn back less than
that. Also many lending desks might chose to make loans in these securities
as part of a broader loan arrangement with other securities or they
could bundle it up with other securities as collateral for other positions:
(Sakurai \& Uchida 2014; Duffie, Scheicher \& Vuillemey 2015; Fuhrer,
Guggenheim \& Schumacher 2016)\footnote{\begin{doublespace}
\noindent Marketable collateral is the exchange of financial assets,
such as stocks and bonds, for a loan between a financial institution
and borrower. To be deemed marketable collateral, assets must be capable
of being sold under normal market conditions with reasonable promptness
at a fair market value. Conditions are based upon actual transactions
on an auction or similarly available daily bid, or ask price market.
\href{https://en.wikipedia.org/wiki/Marketable_collateral}{Marketable Collateral, Wikipedia Link}
\end{doublespace}
}. Given their low contribution to the P\&L, these securities are not
to be placed in the top short ranking, irrespective of the values
of other factors. 

\begin{figure}[h]
\includegraphics[width=17cm]{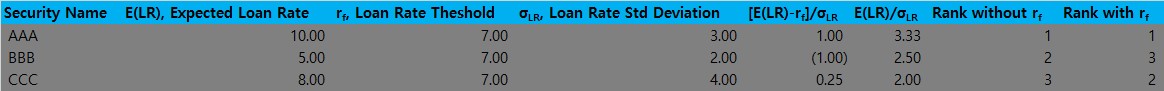}

\caption{\label{fig:Ranking-wwo_Threshold}Ranking with and without Rate Threshold}
\end{figure}

Figure (\ref{fig:Ranking-wwo_Threshold}) can help us to understand
why we need the minimum loan rate threshold and why just the ratio
of return to risk is not enough. The columns from left to right in
Figure (\ref{fig:Ranking-wwo_Threshold}) represent the following
information: 1) security name; 2) expected loan rate; 3) loan rate
threshold; 4) loan rate standard deviation 5) score calculated using
the formula $\left[E\left(LR\right)-r_{f}\right]/\sigma_{LR}$; 6)
score calculated using the formula $E\left(LR\right)/\sigma_{LR}$;
7) ranking of the security using the score calculated with $r_{f}$
; 8) and ranking of the security using the score calculated without
$r_{f}$.

As we see in Figure (\ref{fig:Ranking-wwo_Threshold}), security BBB
with an expected loan rate, $E\left(LR\right)=5.00$, and a loan rate
standard deviation, $\sigma_{LR}$= 2.00, ranks higher than security
CCC with $E\left(LR\right)=8.00$ and $\sigma_{LR}$= 4.00 when we
consider just the ratio of expected loan rate and loan rate standard
deviation, $E\left(LR\right)/\sigma_{LR}$. This issue is fixed and
security CCC is ranked higher than BBB when we consider the excess
loan rate or loan rate premium divided by the loan rate standard deviation,
$\left[E\left(LR\right)-r_{f}\right]/\sigma_{LR}$, to obtain a score. 

We could extend the above short score (Equation: \ref{eq:2}) to include
the short interest, loan availability, days to cover and the loan
balance growth to get the formulations shown in Equations (\ref{eq:3};
\ref{eq:4}; \ref{eq:5}). 
\begin{equation}
\text{Short Score Two}=\left\{ \frac{\xi\left(SI\right)}{\xi\left(LA\right)}\right\} \left\{ \frac{E\left(LR\right)-r_{f}}{\sigma_{LR}}\right\} \label{eq:3}
\end{equation}
 Below, $\xi\left(\right)$ represents the moving average function
applied over the last few (two or three) months. 
\begin{equation}
\text{Short Score Three}=\left\{ \frac{SI}{ADV}\right\} \left\{ \frac{\xi\left(SI\right)}{\xi\left(LA\right)}\right\} \left\{ \frac{E\left(LR\right)-r_{f}}{\sigma_{LR}}\right\} \equiv\left\{ DTC\right\} \left\{ \frac{\xi\left(SI\right)}{\xi\left(LA\right)}\right\} \left\{ \frac{E\left(LR\right)-r_{f}}{\sigma_{LR}}\right\} \label{eq:4}
\end{equation}
Below, $LB_{t}$ is the loan balance at time $t$. $N$ is the number
of days or observations in the time series data.
\begin{equation}
\text{Short Score Four}=\left(\frac{LB_{t}}{LB_{t-N}}\right)\left(DTC\right)\left\{ \frac{\xi\left(SI\right)}{\xi\left(LA\right)}\right\} \left\{ \frac{E\left(LR\right)-r_{f}}{\sigma_{LR}}\right\} \label{eq:5}
\end{equation}

Due to the multiplication of factors, the loan balance growth is better
represented as the ratio of the loan balance at the end and start
of the time period under consideration. We use a ratio instead of
expressing growth in percentage terms, which is more suitable for
a summation. A higher ratio indicates an increase in the loan balance
contribution to a higher score and vice versa, $LBG=LB_{t}/LB_{t-N}$.
We also take the moving average of some variables since without taking
the moving average the ranking would be sensitive to daily changes
and might change quite drastically from day to day. Scores based on
the moving average are more stable. Hence depending on the trading
strategy, whether we are re-balancing the portfolio on a weekly basis
or once in a few months, we can choose either the moving average or
the values on a particular day. 

The scores can also be highly sensitive since they combine the sensitivities
of the individual factors. This requires watching out for outliers
and also having filters and other constraints that can remove the
outliers. Depending on the statistical properties of the variables
under consideration, rules can be established to set bounds on the
extent to which the ratios and hence the scores can change over a
given time period. This will provide alerts when there are huge changes
and aid in managing the sensitivities.

Other extensions could use a SR type score for each variable and combine
it with the other variables. Instead of the risk free rate, we would
need to use a suitable cutoff point for the other variables, such
that when the variable drops below that point, it gets taken out of
the top ranking by having a negative score. We need to pay attention
while combining such ratios since if more than two variables have
a negative value and they are multiplied together, the resultant score
could still be positive. 

When considering any measure that is similar to the SR, the total
variation (or risk) will have a systematic and unsystematic component.
The changes in the variable due to these market or individual sources
will affect the maximum or minimum values that the variable will take
on during short and long term horizons. The effect of the idiosyncratic
or non-idiosyncratic sources on each of the factors could be different.
To isolate these effects we would need to run regressions across historical
data sets to calculate the corresponding coefficients, which can provide
a way to gauge the sensitivities. The challenge would be to create
a proxy for the market specific to the factor under consideration. 

The sources of risk will also vary across different types of crisis.
As we have observed, the global financial crisis affected the financial
sector initially and then spread to other sectors. The COVID-19 pandemic
is having a much more severe impact on the travel and hospitality
sectors. The sources of risk will differ across industry, region,
market capitalization and other security classifications. Pursuing
this avenue, of better understanding the sensitivities from systematic
and unsystematic components including the variations across groups
of securities, also opens the door to many advanced econometric techniques.
\end{doublespace}
\begin{doublespace}

\section{\label{sec:Data-Generation-via}Data Generation via Simulation}
\end{doublespace}

\begin{doublespace}
The bounce basket (and other such baskets) can be created by various
sell side brokers for buy side firms. Having an idea of how such baskets
are created, including the factors used based on the discussion in
Sections (\ref{sec:Securities-Lending-Factors}; \ref{sec:Sharpening-the-Sharpe}),
can lead to meaningful discussions between the two parties that want
to participate in such a scheme. An in-depth comprehension of the
techniques involved can be immensely useful towards maximizing the
benefits of such strategies. The data-set required for an in-house
flavor of the above idea, for risk monitoring or fine tuning the parameters,
can be obtained by any investment firm from the securities lending
desks of their prime brokers (Hildebrand 2007; Melvin \& Taylor 2009;
Jacobs \& Levy 1993)\footnote{\begin{doublespace}
\noindent Prime brokerage is the generic name for a bundled package
of services offered by investment banks, wealth management firms,
and securities dealers to hedge funds which need the ability to borrow
securities and cash in order to be able to invest on a netted basis
and achieve an absolute return. The prime broker provides a centralized
securities clearing facility for the hedge fund so the hedge fund's
collateral requirements are netted across all deals handled by the
prime broker. These two features are advantageous to their clients.
The prime broker benefits by earning fees (\textquotedbl spreads\textquotedbl )
on financing the client's margined long and short cash and security
positions, and by charging, in some cases, fees for clearing and other
services. It also earns money by re-hypothecating the margined portfolios
of the hedge funds currently serviced and charging interest on those
borrowing securities and other investments. Re-hypothecation occurs
when the creditor (a bank or broker-dealer) re-uses the collateral
posted by the debtor (a client such as a hedge fund) to back the broker's
own trades and borrowing. This mechanism also enables leverage in
the securities market. \href{https://en.wikipedia.org/wiki/Prime_brokerage}{Prime Broker, Wikipedia Link}
\end{doublespace}
}.

As noted earlier, given the number of random variables involved (and
hence the complexity of the system), a certain amount of computational
infrastructure would be necessary.\textcolor{black}{{} A typical securities
lending desk can have loan positions on anywhere from a few hundred
to upwards of a few thousand different securities and many years of
historical data. It is therefore, a good complement to build some
intelligence that utilizes the historical time series and calculates
the short score from the corresponding formulae derived in Section
(\ref{sec:Sharpening-the-Sharpe}) without the need for too many ongoing
changes. This can be accomplished by software routines that automatically
run daily using data received from broker firms.}

\textcolor{black}{To demonstrate how this technique would work, we
simulate the historical time series. As opposed to the intermediary,
who would have all the above information, a typical investment firm
is unlikely to have access to the full historical time series. An
investment firm might have the time series of loan rates, prices,
trading volume and availability and hence would have to simulate the
variables for which the historical data is absent, as shown in this
section, to come up with a short score. We create one hundred different
hypothetical securities }and we come up with the time series of all
the variables involved:\textcolor{black}{{} Price, Availability, Short
Interest, Trading Volume, Loan Rate, Alternate Loan Rate. }We model
these variables as Geometric Brownian Motions (GBMs) with uncertainty
introduced via\textcolor{black}{{} sampling from suitable log normal
distributions or by sampling from suitable absolute normal distributions
(}Equations: \ref{eq:6}; \ref{eq:7}\textcolor{black}{). }

\textcolor{black}{Norstad (1999) has a technical discussion of the
normal and log normal distributions. }Hull \& Basu (2016)\textcolor{black}{{}
provide an excellent account of using GBMs to model stock prices and
other time series that are always positive. It is worth noting that
the starting value, mean and standard deviation of the time series
are themselves simulations from other appropriately chosen uniform
distributions (Figure \ref{fig:Simulation-Seed}). Some of the above
variables can be modeled as Poisson distributions or we can simply
consider them as the absolute value of a normal distribution with
appropriately chosen units. As an example, we model the Loan Balance
process as a folded normal distribution or by taking the absolute
value of a normal distribution. The mean and standard deviation of
the loan balance distribution for each security are chosen from another
appropriately chosen uniform distribution.}

A GBM is characterized as below. $S_{it}$ is the stochastic process
that follows a GBM by satisfying the below stochastic differential
equation (Equation: \ref{eq:6}). $S_{i}$ could be the price or another
variable that always takes positive values of the $i^{th}$ security.
$\mu_{S_{i}}$ is the drift and $\sigma_{S_{i}}$ is the volatility.
$W_{t}^{S_{i}}$ is the Weiner Process governing the variable $S_{i}^{th}$
variable. 

\begin{equation}
\text{Geometric Brownian Motion }\equiv\frac{dS_{it}}{S_{it}}=\mu_{S_{i}}dt+\sigma_{S_{i}}dW_{t}^{S_{i}}\label{eq:6}
\end{equation}

Alternately, we could sample the variable values, $S_{it}$, from
an absolute normal distribution with mean, $\mu_{S_{i}}$, and variance,
$\sigma_{S_{i}}^{2}$, as shown in Equation (\ref{eq:7}).

\begin{align}
\text{Alternately},\;S_{it} & \sim\left|N\left(\mu_{S_{i}},\sigma_{S_{i}}^{2}\right)\right|,\;\text{Absolute\;\ Normal\;\ Distribution}\label{eq:7}
\end{align}

The simulation seed is chosen so that the drift and volatility we
get for the variables are similar to what would be observed in practice.
For example in Figure (\ref{fig:Simulation-Seed}), the price and
rate volatility are lower than the volatilities of the availability
and other quantities, which tend to be much higher. The range of the
drift for the quantities, which are share amounts, is also higher
as compared to the drift range of prices and rates.  This ensures
that we are keeping it as close to a realistic setting as possible,
without having access to an actual historical time series. The volatility
and drift, which are proxies for the\textbf{ }standard deviation and
mean for the loan balance process, of the variables for each security
are shown in Figure (\ref{fig:Simulation-Sample-Distributions}).
The length of the simulated time series is a little longer than one
year or around 252 trading days for each security. A sample of the
time series of the variables generated using the simulated drift and
volatility parameters is shown in Figure (\ref{fig:Simulation-Sample-Time}).
The full time series corresponding to the sample shown in the figures
is available upon request.
\end{doublespace}
\begin{doublespace}

\section{\label{sec:Simple-Sample-Ranking}Simple Sample Ranking}
\end{doublespace}

\begin{doublespace}
In Figures (\ref{fig:Short-Scores--Average}; \ref{fig:Short-Scores--First};
\ref{fig:Short-Scores--Last}) we provide three flavors of the short
scores of the securities based on the 60 day moving average of the
variables, the values on the first day and the values on the last
day of the sample correspondingly. Under each of the three flavors,
we consider all four formulations of the short scores given in Equations
(\ref{eq:2}; \ref{eq:3}; \ref{eq:4}; \ref{eq:5}) from Section
(\ref{sec:Sharpening-the-Sharpe}).

Comparing the scores on the first day and last day should give us
an idea of how the scores, and hence the ranking, can change over
the duration of a few months. Also, availability can be zero for certain
securities, leading to a error in the score calculation when we divide
by zero, which is avoided in many cases by using a moving average.
Since our scores are based on a simulation we see some low loan rates,
low availability and high short interest numbers. In general, low
availability and high short interest will be reflected as higher loan
rates. Similarly high loan rates, high availability and low short
interest rarely occur in practice. The days to cover generally does
not include the moving average of the short interest, but we use the
moving average only for the flavor in Figure (\ref{fig:Short-Scores--Average}).

The initial ranking of top shorts is ideally done over a large universe
and the final basket will need around a couple of dozen stocks. Hence,
filters need to be chosen to narrow down the original list. We can
remove a certain percentage of securities from the bottom (say 20
percent) from this ranked set of securities and additional filters
can be applied (sector, market capitalization, etc.) to further narrow
down the constituents of the basket. Some details about how filters
can be applied are provided in the explanation of the factors (Section
\ref{sec:Securities-Lending-Factors}) and the selection methodology
above. Some of the markets on which empirical tests using real data
were tried, at various investment firms we have been involved with
in the past, include Japan, Hong Kong, Taiwan, Korea, Singapore, Thailand,
Indonesia and Malaysia. Custom baskets for each market or even for
individual sectors can be created based on the same selection model.
\end{doublespace}
\begin{doublespace}

\section{\label{sec:Portfolio-Construction}Portfolio Construction and Robustness}
\end{doublespace}

\begin{doublespace}
Once a top shorts ranking has been created and the securities are
ranked accordingly, we can pick a certain number of securities from
the ranking above a threshold value of the short score. We provide
one technique to construct a portfolio, even though most standard
rules used to construct portfolios can be applied given that we have
a way to rank and select securities (Elton, Gruber, Brown \& Goetzmann
2009; Bodie, Kane \& Marcus 2013). The securities in the basket can
simply be weighted based on a combination of their top shorts ranking
and market capitalization, with a maximum weight of say 10\% for any
single security to ensure sufficient diversification. As discussed
in Kashyap (2019) the number of securities we pick needs to account
for the costs of trade executions.

One particular way to weight the securities in a portfolio is shown
in Equations (\ref{eq:Portfolio-Weights}; \ref{eq:Portfolio-Weights-Two}).
The intuition behind the weighting is that the securities that have
a higher short score are weighted higher in the overall portfolio.
The market capitalization or average trading volume can also be included
in the short score using the mechanism we have shown in Section (\ref{sec:Sharpening-the-Sharpe}).
The weight of a particular security will be higher proportional to
the extent by which the short score and the trading volume exceed
the corresponding metrics for the other constituent securities.

The weights of individual securities in the portfolio, $u_{i}$ are
given below. $SS_{i}$ is the short score of security $i$ and there
are a total of $M$ securities selected for the portfolio.

\begin{equation}
u_{i}=\frac{SS_{i}}{\sum_{i=1}^{M}SS_{i}}\label{eq:Portfolio-Weights}
\end{equation}
\begin{equation}
\text{Also, }\sum_{i=1}^{M}u_{i}=1\label{eq:Portfolio-Weights-Two}
\end{equation}

We next describe some mechanisms that can bring in a certain amount
of robustness to the constructed portfolio. One simple way to make
the portfolio more robust would be to measure the changes in the rankings
of the variables over time and penalize the securities with greater
movement in their rankings. This would mean that securities with greater
movement in the scores and hence the rankings would have a lower weight.
Also, we could change the security weights only when the change in
the scores and hence the rankings are beyond certain thresholds. This
would ensure that minor changes would not require rebalancing the
portfolio. 

A more sophisticated way to weight the securities can be based on
the variance of the short score of the securities over a historical
time period. Kashyap (2016) provides an example of such a weighting,
which has an intuitive and practical appeal since the time series
with a higher variance is set a lower weight and hence represented
accordingly in the resultant portfolio. Similar to the description
in Section (\ref{sec:Securities-Lending-Factors}) we could apply
filters based on individual variables to ensure that outliers are
not added into the portfolio. 

Any portfolio constructed using scores based on an SR type metric
(Section \ref{sec:Sharpening-the-Sharpe}) will be sensitive to the
two sources of variation in the factors, systematic and unsystematic,
which will influence the total return that can be earned on the portfolio
across different time durations. In a portfolio constructed purely
using the risk and return of security prices, Markowitz diversification
assumptions will ensure that we can minimize the idiosyncratic risk
as we add more securities to the portfolio. But for some of the other
factors we are using, the diversification benefits would need to be
tested more thoroughly since there could be heavy skew towards the
unsystematic components. The number of securities required to bring
in diversification benefits could be significantly different as well.

Sector related considerations, in terms of overweighting or underweighting
a particular sector, can be applied. These sector considerations would
depend on the type of crisis that is unfolding, its causes and the
effect it is having. As we discussed earlier in Section (\ref{sec:Sharpening-the-Sharpe}),
during the COVID-19 crisis travel and hospitality sectors have taken
a greater hit. Hence as the wider market recovers, the sectors that
have been more badly affected might take longer to bounce back and
it would be judicious to underweight the corresponding sectors. It
is possible that, once the troubled sectors start to recover, they
might warrant a greater weighting. Some subjective approaches might
be necessary. If we need to follow a completely data driven approach,
measuring sector specific effects due to the crisis and performing
an allocation to different sectors based on their performance before
the event would be necessary.

Securities with negative events in terms of credit rating downgrades,
earnings downgrades, ongoing litigation or the commencement of new
litigation, failure to close a major deal, loss of a significant customer
etc., are removed on a case by case basis. The source for this information
can be market data vendors or credit rating companies. This step is
included as a fail-safe procedure since the analyst rating filters
and short factors we have in place would remove most of these securities. 

Baskets can be constructed using this same concept by utilizing data
from different historical periods when there is an upward rally. We
have observed, from our earlier experience with actual empirical data,
that the bounce basket has been found to outperform the broader market.
The portfolio construction and portfolio performance are not shown
in the numerical examples since the data is fictitious, but these
steps follow quite easily based on the techniques we have provided.
\end{doublespace}
\begin{doublespace}

\section{\label{sec:Taming-the-Volatility}Taming the Volatility Skew}
\end{doublespace}

\begin{doublespace}
A key improvement in the bounce basket\footnote{\noindent All the strategies discussed in Kashyap (2019) would benefit
from this discussion.} would be some method to factor in the effect of changes in the upward
or downward movements of a variable over time. Such directional changes
could be either beneficial or detrimental to our trading position,
depending on whether we are long or short. This would be especially
important for the purpose of creating any ranking, which is central
to the bounce basket, since we would need to know how steady the changes
in a variable are in terms of whether it is trending upwards or downwards.

One of the main drawbacks of volatility is that it fails to capture
the effect of changes in the direction in the time series of any variable.
This is because both rising or falling prices still have volatility
and if we are long a security, upwardly mobile prices could get penalized
in any portfolio management decisions. Figures (\ref{fig:Limitations-of-Volatility-1};
\ref{fig:Limitations-of-Volatility-2}; \ref{fig:Limitations-of-Volatility-3})
illustrate this limitation of using volatility. In all three figures,
we compare two variables with different time paths but which offer
the same overall return over the time period under consideration. 

In Figure (\ref{fig:Limitations-of-Volatility-1}), variable one and
variable two both have a steady upward path (not a single down movement),
but variable two, which has higher volatility, ends up having a lower
return to volatility ratio (return to risk ratio, if volatility is
used to measure risk) and could get underrepresented in a portfolio
in comparison with variable one. This is clearly a sub-optimal outcome.
Likewise, in Figure (\ref{fig:Limitations-of-Volatility-2}), variable
two has lower volatility, but has falling prices for some time intervals,
when compared to variable one which does not have a single fall in
price. Lastly, in Figure (\ref{fig:Limitations-of-Volatility-3}),
variable one is falling more steadily than variable two, but variable
one has greater volatility and hence can be seen as better than variable
two from a return to volatility perspective.

These examples show that volatility is not the most conducive metric
to represent risk. Clearly, these are simplified examples meant to
illustrate the limitations of volatility. When we use average returns,
either geometric\footnote{In mathematics, the geometric mean is a mean or average, which indicates
the central tendency or typical value of a set of numbers by using
the product of their values (as opposed to the arithmetic mean which
uses their sum). \href{https://en.wikipedia.org/wiki/Geometric_mean}{Geometric Mean, Wikipedia Link}} or arithmetic\footnote{In mathematics and statistics, the arithmetic mean, or simply the
mean or the average (when the context is clear), is the sum of a collection
of numbers divided by the count of numbers in the collection. \href{https://en.wikipedia.org/wiki/Arithmetic_mean}{Arithmetic Mean, Wikipedia Link}}, we do consider the path taken by the variable to a certain extent.
But whenever we form expectations of risk and return, without considering
the up and down movements more carefully, and use risk-return based
measures to form portfolios, the path agnostic nature of volatility
can be costly.

As an alternative to using volatility, we are working on creating
a specially developed factor called the time weighted arrival deviation
(Kashyap 2017a). This factor can be used as a metric to measure the
path taken by any variable to arrive at the current value, over the
last few months or last few time periods. To articulate the intuition,
any security that had steady upward growth, over a particular time
period, is ranked higher than a security with similar growth, over
the same time period, but with more ups and downs in the path, or
changes in direction. Similarly, any security that had steady downward
growth is ranked lower than a security with similar downward growth
but with more changes in direction in the path. To summarize, securities
are penalized for having ups and downs in their price process. But
unidirectional jumps in the direction of intended movement suffer
no such penalty, since if our position is long (or short) then we
want the price to go up (or down).
\end{doublespace}
\begin{doublespace}

\section{\label{sec:Conclusion}Conclusion and Possibilities for Improvement}
\end{doublespace}

\begin{doublespace}
We have discussed in detail a trading strategy, called the bounce
basket, for someone to express a bullish view on the market by allowing
them to take long positions on securities that would benefit the most
from a rally in the markets. This investment idea could be a stock
selection mechanism for active investors given its inherent focus
on the selection of securities that are expected to outperform the
market or a corresponding securities index. We have come up with a
possible behavioral bias that might be prevalent in the financial
markets (possibly also present in many other areas of life), which
we have termed the rebound effect. The trading idea we have outlined
could be a way to overcome behavioral biases that investors might
harbor, which might lead investors to trade in an unrestrained manner,
after experiencing losses due to market crashes.

We have illustrated the components of creating our strategy including
the mechanics of constructing the portfolio. Using simulated data,
we have given a few flavors of creating a top shorts ranking of securities.
We have considered many pointers on how our idea can be made more
robust including the practical aspects one needs to pay attention
to while implementing our investment strategy. The variables we have
discussed in detail are the more commonly available ones, though we
have mentioned other supplementary variables which can be harder to
acquire but can improve the decision process. Our discussion holds
many lessons for developing other investment ideas and putting them
into action. We point the readers to a comprehensive collection of
trading strategies and market color information in Kashyap (2019).

Future investigations regarding the rebound effect should concentrate
on verifying, under different scenarios and assumptions, whether such
an effect actually exists with regards to financial investments. Investor
reactions after making losses (human behavior before and after hardship)
can be a very interesting avenue for exploration. Some possibilities
are to look at how soon investor intentions are to return back to
making investments, changes to the size of investments, any alterations
to their risk appetites and desired asset classes. Though, as we have
outlined earlier, even if the rebound effect is not significant in
the financial markets or across broad categories of investors, our
strategy can be an excellent security selection tool when markets
are expected to trend upwards. 

We have outlined a key limitation of using volatility, as a measure
of risk, and chronicled our initial efforts to arrive at an alternate
metric that is more sensitive to the path taken by variables. Further
efforts in this line of inquiry can be challenging yet quite rewarding.
Much work can be done in terms of investigating the significance and
variation of the factors over time and across geographical locations.
We have taken care of ensure that the simulations we have run to create
the security ranking mirror real datasets that we have worked with
in the past. The portfolio construction and portfolio performance
are not shown in the numerical examples since the data is fictitious,
but these steps follow quite easily based on the techniques we have
provided. The exact formulae derived in Sections (\ref{sec:Sharpening-the-Sharpe};
\ref{sec:Portfolio-Construction}) can be immediately applied by investors
to implement these strategies in a straight forward manner.

The ongoing state of affairs, due to the COVID-19 pandemic, renders
an ideal scenario for implementing the strategy we have developed.
Stock markets globally have been affected by the pandemic, but as
life returns to normal and markets recover, careful selection of securities
based on our strategy can lead to outsized returns. Any crisis or
pandemic is bound to flood the media and research outlets, and hence
many aspects of our lives, with negative sentiment. During times of
such adversity, it is not uncommon to be bombarded with news that
seems to adversely affect everything around us including the financial
markets.

Our attempt is to bring some optimism into the present (or any?) distressed
atmosphere by drawing attention to the good things that await us once
we emerge from any calamity. As has been told many times in the past,
the night seems darkest before dawn. Hence, as the crisis seems to
rage on at its fiercest, our desire is that this paper can be a tiny
beacon of radiance and a glimmer of hope that will lead to resurgence
in the financial markets and other aspects of life affected by this
fight against the COVID-19 infectious disease. The many benefits of
holding a positive attitude need to be studied further, but embracing
such an outlook can possibly cause little harm and might give us the
strength to prevail against overwhelming odds.
\end{doublespace}
\begin{doublespace}

\section{References }
\end{doublespace}
\begin{enumerate}
\begin{doublespace}
\item Abbink, K., Irlenbusch, B., \& Renner, E. (2000). The moonlighting
game: An experimental study on reciprocity and retribution. Journal
of Economic Behavior \& Organization, 42(2), 265-277.
\item Afshar, T., Arabian, G., \& Zomorrodian, R. (2007). Stock return,
consumer confidence, purchasing managers index and economic fluctuations.
Journal of Business \& Economics Research (JBER), 5(8).
\item Ammann, M., Kessler, S., \& Tobler, J. (2006). Analyzing active investment
strategies. The Journal of Portfolio Management, 33(1), 56-67.
\item Asquith, P., Pathak, P. A., \& Ritter, J. R. (2005). Short interest,
institutional ownership, and stock returns. Journal of Financial Economics,
78(2), 243-276.
\item Baek, S., Mohanty, S. K., \& Glambosky, M. (2020). COVID-19 and stock
market volatility: An industry level analysis. Finance Research Letters,
37, 101748.
\item Baker, M., \& Wurgler, J. (2007). Investor sentiment in the stock
market. Journal of economic perspectives, 21(2), 129-152.
\item Baker, S. R., Bloom, N., Davis, S. J., Kost, K., Sammon, M., \& Viratyosin,
T. (2020). The unprecedented stock market reaction to COVID-19. The
Review of Asset Pricing Studies, 10(4), 742-758.
\item Bakshi, G., \& Chen, Z. (2005). Stock valuation in dynamic economies.
Journal of Financial Markets, 8(2), 111-151.
\item Balvers, R., Wu, Y., \& Gilliland, E. (2000). Mean reversion across
national stock markets and parametric contrarian investment strategies.
The Journal of Finance, 55(2), 745-772.
\item Barber, B., Lehavy, R., McNichols, M., \& Trueman, B. (2001). Can
investors profit from the prophets? Security analyst recommendations
and stock returns. The Journal of Finance, 56(2), 531-563.
\item Barber, B., Lehavy, R., McNichols, M., \& Trueman, B. (2003). Reassessing
the returns to analysts' stock recommendations. Financial Analysts
Journal, 59(2), 88-96.
\item Barber, L. L., \& Cooper, M. L. (2014). Rebound sex: Sexual motives
and behaviors following a relationship breakup. Archives of Sexual
Behavior, 43(2), 251-265.
\item Berg, J., Dickhaut, J., \& McCabe, K. (1995). Trust, reciprocity,
and social history. Games and economic behavior, 10(1), 122-142.
\item Berkhout, P. H., Muskens, J. C., \& Velthuijsen, J. W. (2000). Defining
the rebound effect. Energy policy, 28(6-7), 425-432. 
\item Buchan, N. R., Croson, R. T., \& Solnick, S. (2008). Trust and gender:
An examination of behavior and beliefs in the Investment Game. Journal
of Economic Behavior \& Organization, 68(3-4), 466-476.
\item Bodie, Z., Kane, A., \& Marcus, A. J. (2013). Investments and portfolio
management. McGraw Hill Education (India) Private Limited.
\item Bris, A., Goetzmann, W. N., \& Zhu, N. (2007). Efficiency and the
bear: Short sales and markets around the world. The Journal of Finance,
62(3), 1029-1079.
\item Brown, N. C., Wei, K. D., \& Wermers, R. (2013). Analyst recommendations,
mutual fund herding, and overreaction in stock prices. Management
Science, 60(1), 1-20.
\item Browne, S. (2000). Risk-constrained dynamic active portfolio management.
Management Science, 46(9), 1188-1199.
\item Brumbaugh, C. C., \& Fraley, R. C. (2015). Too fast, too soon? An
empirical investigation into rebound relationships. Journal of Social
and Personal Relationships, 32(1), 99-118.
\item Charness, G., Cobo-Reyes, R., \& Jiménez, N. (2008). An investment
game with third-party intervention. Journal of Economic Behavior \&
Organization, 68(1), 18-28.
\item Choudhry, T. (1996). Stock market volatility and the crash of 1987:
evidence from six emerging markets. Journal of International money
and Finance, 15(6), 969-981.
\item Cox, J. C., \& Leland, H. E. (2000). On dynamic investment strategies.
Journal of Economic Dynamics and Control, 24(11-12), 1859-1880.
\item De Bondt, W. F., \& Thaler, R. (1985). Does the stock market overreact?.
The Journal of finance, 40(3), 793-805.
\item Dechow, P. M., Hutton, A. P., Meulbroek, L., \& Sloan, R. G. (2001).
Short-sellers, fundamental analysis, and stock returns. Journal of
Financial Economics, 61(1), 77-106.
\item Duffie, D., Scheicher, M., \& Vuillemey, G. (2015). Central clearing
and collateral demand. Journal of Financial Economics, 116(2), 237-256.
\item Eichengreen, B., \& Irwin, D. A. (2010). The slide to protectionism
in the great depr - ession: who succumbed and why?. The Journal of
Economic History, 70(4), 871-897.
\item Elton, E. J., Gruber, M. J., Brown, S. J., \& Goetzmann, W. N. (2009).
Modern portfolio theory and investment analysis. John Wiley \& Sons.
\item Frederick, S. (2005). Cognitive reflection and decision making. Journal
of Economic perspectives, 19(4), 25-42.
\item Fudenberg, D., \& Tirole, J. (1991). Game theory mit press. Cambridge,
MA, 86.
\item Fuhrer, L., Guggenheim, B., \& Schumacher, S. (2016). Re‐Use of Collateral
in the Repo Market. Journal of Money, Credit and Banking, 48(6), 1169-1193.
\item Gibbons, R. S. (1992). Game theory for applied economists. Princeton
University Press.
\item Gillingham, K., Kotchen, M. J., Rapson, D. S., \& Wagner, G. (2013).
The rebound effect is overplayed. Nature, 493(7433), 475-476.
\item Guo, Y. R., Cao, Q. D., Hong, Z. S., Tan, Y. Y., Chen, S. D., Jin,
H. J., ... \& Yan, Y. (2020). The origin, transmission and clinical
therapies on coronavirus disease 2019 (COVID-19) outbreak--an update
on the status. Military Medical Research, 7(1), 1-10.
\item Granados, J. A. T., \& Roux, A. V. D. (2009). Life and death during
the Great Depression. Proceedings of the national academy of sciences,
106(41), 17290-17295. 
\item Granger, C. W. (1969). Investigating causal relations by econometric
models and cross-spectral methods. Econometrica: journal of the Econometric
Society, 424-438.
\item Hanspal, T., Weber, A., \& Wohlfart, J. (2020). Exposure to the COVID-19
stock market crash and its effect on household expectations. Review
of Economics and Statistics, 1-45.
\item Hildebrand, P. M. (2007). Hedge funds and prime broker dealers: steps
towards a “best practice proposal”. FSR FINANCIAL, 67.
\item Hull, J. C., \& Basu, S. (2016). Options, futures, and other derivatives.
Pearson Education India.
\item Ismail, S. (2014). Exponential Organizations: Why new organizations
are ten times better, faster, and cheaper than yours (and what to
do about it). Diversion Books.
\item Jacobs, B. I., \& Levy, K. N. (1993). Long/short equity investing.
Journal of Portfolio Management, 20(1), 52.
\item Jang, H., \& Sul, W. (2002). The Asian financial crisis and the co-movement
of Asian stock markets. Journal of Asian Economics, 13(1), 94-104.
\item Javelle, E., \& Raoult, D. (2020). COVID-19 pandemic more than a century
after the Spanish flu. The Lancet Infectious Diseases.
\item Kashyap, R (2016). Securities Lending Strategies: Exclusive Auction
Bids. Working Paper.
\item Kashyap, R (2017a). Time Weighted Arrival Deviation: Moving from the
Efficient Frontier towards the Final Frontier of Investments. Working
Paper.
\item Kashyap, R (2017b). Securities Lending Strategies, TBR and TBR (Theoretical
Borrow Rate and Thoughts Beyond Rates). Working Paper.
\item Kashyap, R. (2019). Concepts, Components, and Collections of Trading
Strategies and Market Color. The Journal of Wealth Management, 22(3),
115-128.
\item Kashyap, R. (2021). Artificial Intelligence: A Child’s Play. Technological
Forecasting and Social Change, 166, 120555.
\item Khan, S., \& Park, K. W. K. (2009). Contagion in the stock markets:
The Asian financial crisis revisited. Journal of Asian Economics,
20(5), 561-569.
\item Kindleberger, C. P., \& Aliber, R. Z. (2011). Manias, panics and crashes:
a history of financial crises. Palgrave Macmillan. 
\item Koenig, E. F. (2002). Using the purchasing managers’ index to assess
the economy’s strength and the likely direction of monetary policy.
Federal Reserve Bank of Dallas Economic and Financial Policy Review,
1(6), 1-14.
\item Kudryavtsev, A., Cohen, G., \& Hon-Snir, S. (2013). 'Rational'or'Intuitive':
Are Behavioral Biases Correlated Across Stock Market Investors?. Contemporary
economics, 7(2), 31-53.
\item Lee, S. B., \& Kim, K. J. (1993). Does the October 1987 crash strengthen
the co‐movements among national stock markets?. Review of Financial
Economics, 3(1), 89-102.
\item Loewenstein, G. (2000). Emotions in economic theory and economic behavior.
American economic review, 90(2), 426-432.
\item Malkiel, B. G. (2003). Passive investment strategies and efficient
markets. European Financial Management, 9(1), 1-10.
\item Mazur, M., Dang, M., \& Vega, M. (2021). COVID-19 and the march 2020
stock market crash. Evidence from S\&P1500. Finance Research Letters,
38, 101690.
\item Melvin, M., \& Taylor, M. P. (2009). The crisis in the foreign exchange
market. Journal of International Money and Finance, 28(8), 1317-1330.
\item Menkhoff, L., \& Nikiforow, M. (2009). Professionals’ endorsement
of behavioral finance: Does it impact their perception of markets
and themselves?. Journal of Economic Behavior \& Organization, 71(2),
318-329.
\item Morens, D. M., Breman, J. G., Calisher, C. H., Doherty, P. C., Hahn,
B. H., Keusch, G. T., ... \& Taubenberger, J. K. (2020). The origin
of COVID-19 and why it matters. The American journal of tropical medicine
and hygiene, 103(3), 955-959.
\item Nikkinen, J., Piljak, V., \& Äijö, J. (2012). Baltic stock markets
and the financial crisis of 2008--2009. Research in International
Business and Finance, 26(3), 398-409.
\item Norstad, J. (1999). The normal and lognormal distributions.
\item Odekon, M. (2015). Booms and Busts: An Encyclopedia of Economic History
from the First Stock Market Crash of 1792 to the Current Global Economic
Crisis. Routledge.
\item Oechssler, J., Roider, A., \& Schmitz, P. W. (2009). Cognitive abilities
and behavioral biases. Journal of Economic Behavior \& Organization,
72(1), 147-152.
\item Ohanian, L. E. (2009). What--or who--started the great depression?.
Journal of Economic Theory, 144(6), 2310-2335. - Herbert Hoover.
\item Payne, J. W., Laughhunn, D. J., \& Crum, R. (1980). Translation of
gambles and aspiration level effects in risky choice behavior. Management
Science, 26(10), 1039-1060. 
\item Pelaez, R. F. (2003). A reassessment of the purchasing managers' index.
Business Economics, 38(4), 35-42.
\item Petersen, E., Koopmans, M., Go, U., Hamer, D. H., Petrosillo, N.,
Castelli, F., ... \& Simonsen, L. (2020). Comparing SARS-CoV-2 with
SARS-CoV and influenza pandemics. The Lancet infectious diseases.
\item Phan, D. H. B., \& Narayan, P. K. (2020). Country responses and the
reaction of the stock market to COVID-19---A preliminary exposition.
Emerging Markets Finance and Trade, 56(10), 2138-2150.
\item Pompian, M. M. (2011). Behavioral finance and wealth management: how
to build investment strategies that account for investor biases (Vol.
667). John Wiley \& Sons.
\item Ranaldo, A., \& Haeberle, R. (2008). Wolf in sheep's clothing: the
active investment strategies behind index performance. European Financial
Management, 14(1), 55-81.
\item Reinhart, C. M., \& Rogoff, K. S. (2009). This time is different:
Eight centuries of financial folly. princeton university press.
\item Richardson, G., \& Troost, W. (2009). Monetary intervention mitigated
banking panics during the great depression: quasi-experimental evidence
from a federal reserve district border, 1929--1933. Journal of Political
Economy, 117(6), 1031-1073.
\item Roll, R. (1988). The international crash of October 1987. Financial
analysts journal, 44(5), 19-35.
\item Ross, S. A., Westerfield, R., Jaffe, J. F., \& Jordan, B. D. (2009).
Corporate finance: Core principles \& applications. McGraw-Hill Irwin. 
\item Sakurai, Y., \& Uchida, Y. (2014). Rehypothecation dilemma: Impact
of collateral rehypothecation on derivative prices under bilateral
counterparty credit risk. Journal of Banking \& Finance, 48, 361-373.
\item Schwert, G. W. (1990). Stock volatility and the crash of’87. The review
of financial studies, 3(1), 77-102.
\item Sharpe, W. F. (1966). Mutual fund performance. The Journal of business,
39(1), 119-138.
\item Sharpe, W. F. (1994). The sharpe ratio. Journal of portfolio management,
21(1), 49-58.
\item Shereen, M. A., Khan, S., Kazmi, A., Bashir, N., \& Siddique, R. (2020).
COVID-19 infection: Origin, transmission, and characteristics of human
coronaviruses. Journal of advanced research, 24, 91-98.
\item Sorrell, S., \& Dimitropoulos, J. (2008). The rebound effect: Microeconomic
definitions, limitations and extensions. Ecological Economics, 65(3),
636-649.
\item Spielmann, S. S., Macdonald, G., \& Wilson, A. E. (2009). On the rebound:
Focusing on someone new helps anxiously attached individuals let go
of ex-partners. Personality and Social Psychology Bulletin, 35(10),
1382-1394.
\item Summers, L. H. (1986). Does the stock market rationally reflect fundamental
values?. The Journal of Finance, 41(3), 591-601.
\item Thomas, G., \& Morgan-Witts, M. (2014). The Day the Bubble Burst:
A Social History of the Wall Street Crash of 1929. Open Road Media.
\item Tsuchiya, Y. (2012). Is the Purchasing Managers’ Index useful for
assessing the economy’s strength? A directional analysis. Economics
Bulletin, 32(2).
\item Tversky, A., \& Kahneman, D. (1983). Extensional versus intuitive
reasoning: The conjunction fallacy in probability judgment. Psychological
review, 90(4), 293.
\item van den Bergh, J. C., \& Gowdy, J. M. (2009). A group selection perspective
on economic behavior, institutions and organizations. Journal of Economic
Behavior \& Organization, 72(1), 1-20.
\item Wolfinger, N. H. (2007). Does the rebound effect exist? Time to remarriage
and subsequent union stability. Journal of Divorce \& Remarriage,
46(3-4), 9-20.
\item Zhang, T., Wu, Q., \& Zhang, Z. (2020). Probable pangolin origin of
SARS-CoV-2 associated with the COVID-19 outbreak. Current biology,
30(7), 1346-1351.
\item Zhang, X., Chen, X., Zhang, Z., Roy, A., \& Shen, Y. (2020). Strategies
to trace back the origin of COVID-19. Journal of Infection, 80(6),
e39-e40.
\item Zhou, P., Yang, X. L., Wang, X. G., Hu, B., Zhang, L., Zhang, W.,
... \& Shi, Z. L. (2020). A pneumonia outbreak associated with a new
coronavirus of probable bat origin. nature, 579(7798), 270-273.
\end{doublespace}
\end{enumerate}
\begin{doublespace}

\section{\label{sec:Appendix-of-Figures}Appendix of Figures}
\end{doublespace}

\begin{doublespace}
For each of the figures in this appendix, detailed explanations are
provided in the main body of the text to help facilitate better understanding.
Below, we provide supplementary descriptions for each figure.
\end{doublespace}
\begin{doublespace}

\subsection{Figures for Section (\ref{sec:Data-Generation-via}), Data Generation
via Simulation}
\end{doublespace}

\begin{figure}[h]
\includegraphics[width=12cm]{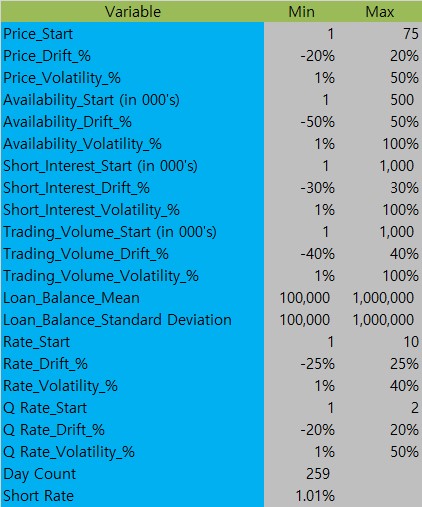}

\caption{Simulation Seed\label{fig:Simulation-Seed}}
\end{figure}

\begin{doublespace}
In Figure (\ref{fig:Simulation-Seed}) we show the minimum and maximum
values that act as the inputs to a uniform distribution. The random
samples chosen from the uniform distribution act as the starting value,
the drift and volatility for the GBMs, corresponding to the one hundred
hypothetical securities that are under consideration, for each of
the following variables: stock price, availability, short interest,
trading volume, loan balance, loan rate and the alternate loan rate.

\begin{figure}[h]
\includegraphics[width=18cm]{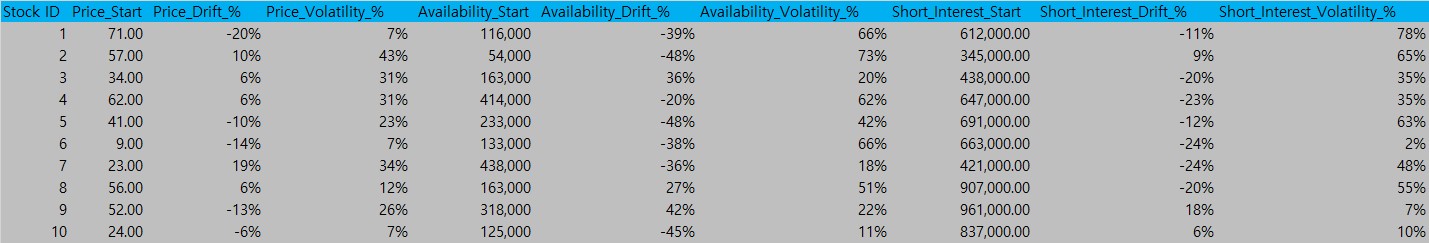}

\includegraphics[width=18cm]{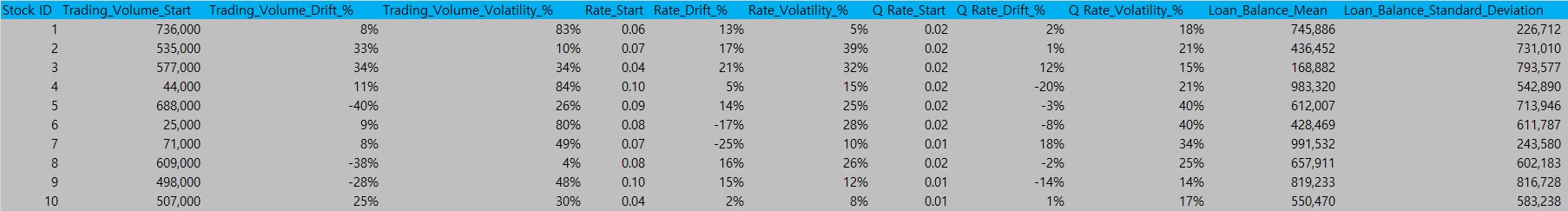}

\caption{Simulation Sample Distributions\label{fig:Simulation-Sample-Distributions}}
\end{figure}

In Figure (\ref{fig:Simulation-Sample-Distributions}) we show some
samples for ten hypothetical securities, out of the total one hundred,
drawn from a uniform distribution with range specified according to
values in Figure (\ref{fig:Simulation-Seed}). The values chosen from
the uniform distribution, which are shown in Figure (\ref{fig:Simulation-Sample-Distributions}),
act as the starting value, the drift and volatility for the GBMs corresponding
to the stock price, availability, short interest, trading volume,
loan balance, loan rate and the alternate loan rate. 

\begin{figure}[h]
\includegraphics[width=12cm]{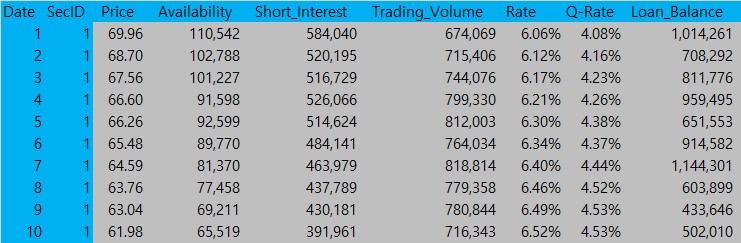}

\includegraphics[width=12cm]{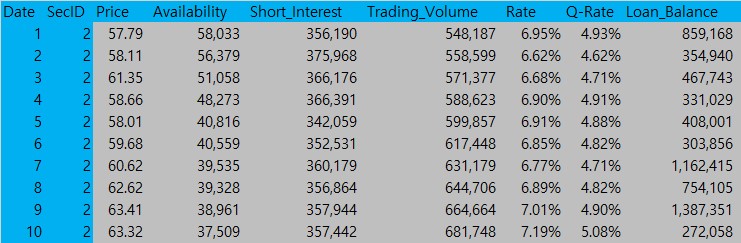}

\caption{Simulation Sample Time Series\label{fig:Simulation-Sample-Time}}
\end{figure}

In Figure (\ref{fig:Simulation-Sample-Time}) we show some sample
time series values created using the corresponding GBMs for two securities
for ten days for the following variables: stock price, availability,
short interest, trading volume, loan balance, loan rate and the alternate
loan rate. The starting value, drift and volatility for the corresponding
GBMs are specified according to the values in Figure (\ref{fig:Simulation-Sample-Distributions}).

\end{doublespace}
\begin{doublespace}

\subsection{Figures for Section (\ref{sec:Simple-Sample-Ranking}), Simple Sample
Ranking}
\end{doublespace}

\begin{figure}[h]
\includegraphics[width=17cm]{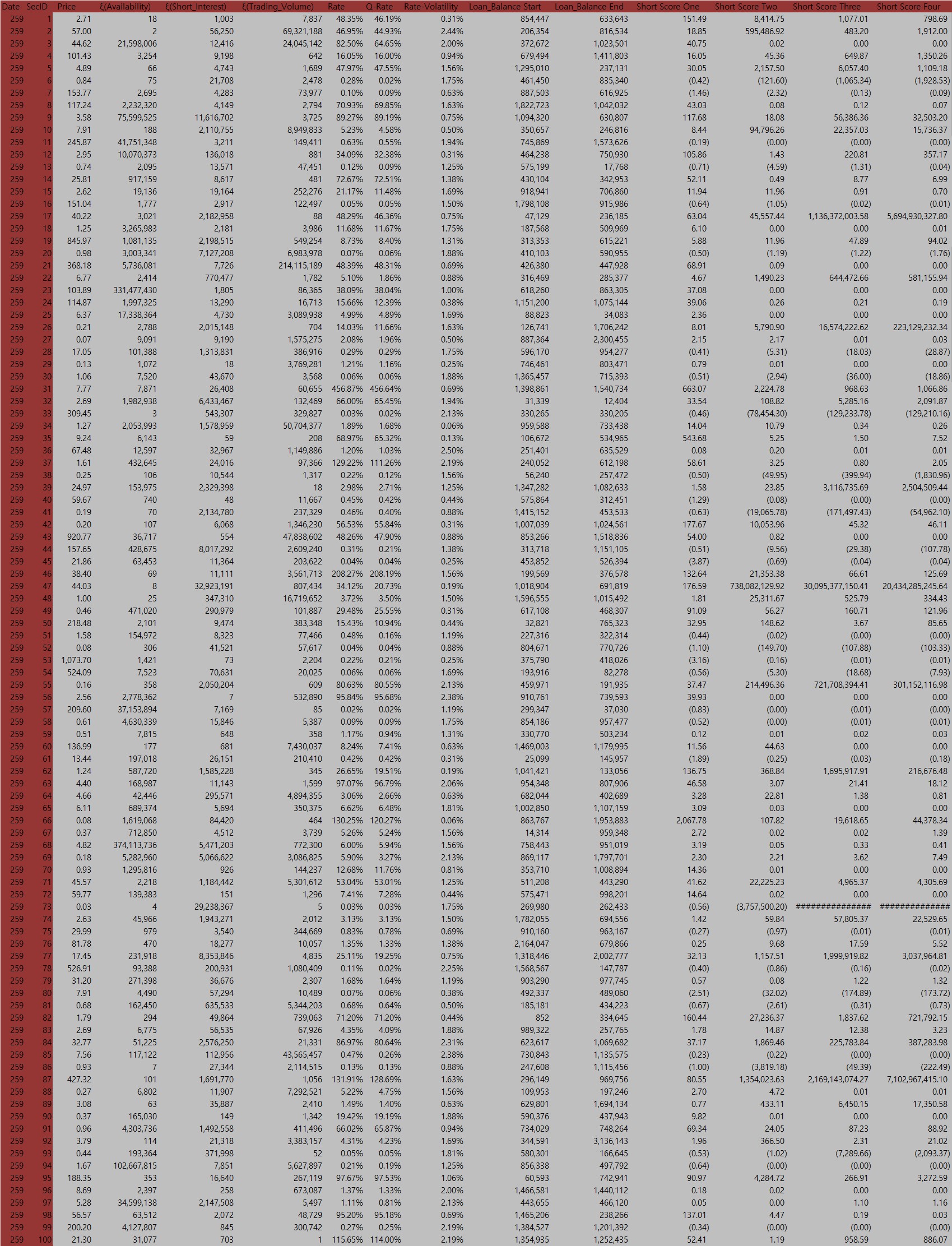}

\caption{\label{fig:Short-Scores--Average}Short Scores - 60 Day Moving Average}
\end{figure}

\begin{doublespace}
In Figure (\ref{fig:Short-Scores--Average}) we provide short scores
for the securities based on the 60 day moving average of some of the
variables. The columns represent the following information respectively:
date, security ID, price, 60 day moving average of the availability,
60 day moving average of the short interest, 60 day moving average
of the trading volume, loan rate, alternate loan rate, rate volatility,
loan balance at the start of the time series sample or on the first
day, loan balance at the end of the time series sample or on the last
day, short score one (Eq: \ref{eq:2}), short score two (Eq: \ref{eq:3}),
short score three (Eq: \ref{eq:4}) and short score four (Eq: \ref{eq:5}).
The values in the table correspond to the date specified in the first
column.

\begin{figure}[h]
\includegraphics[width=17cm]{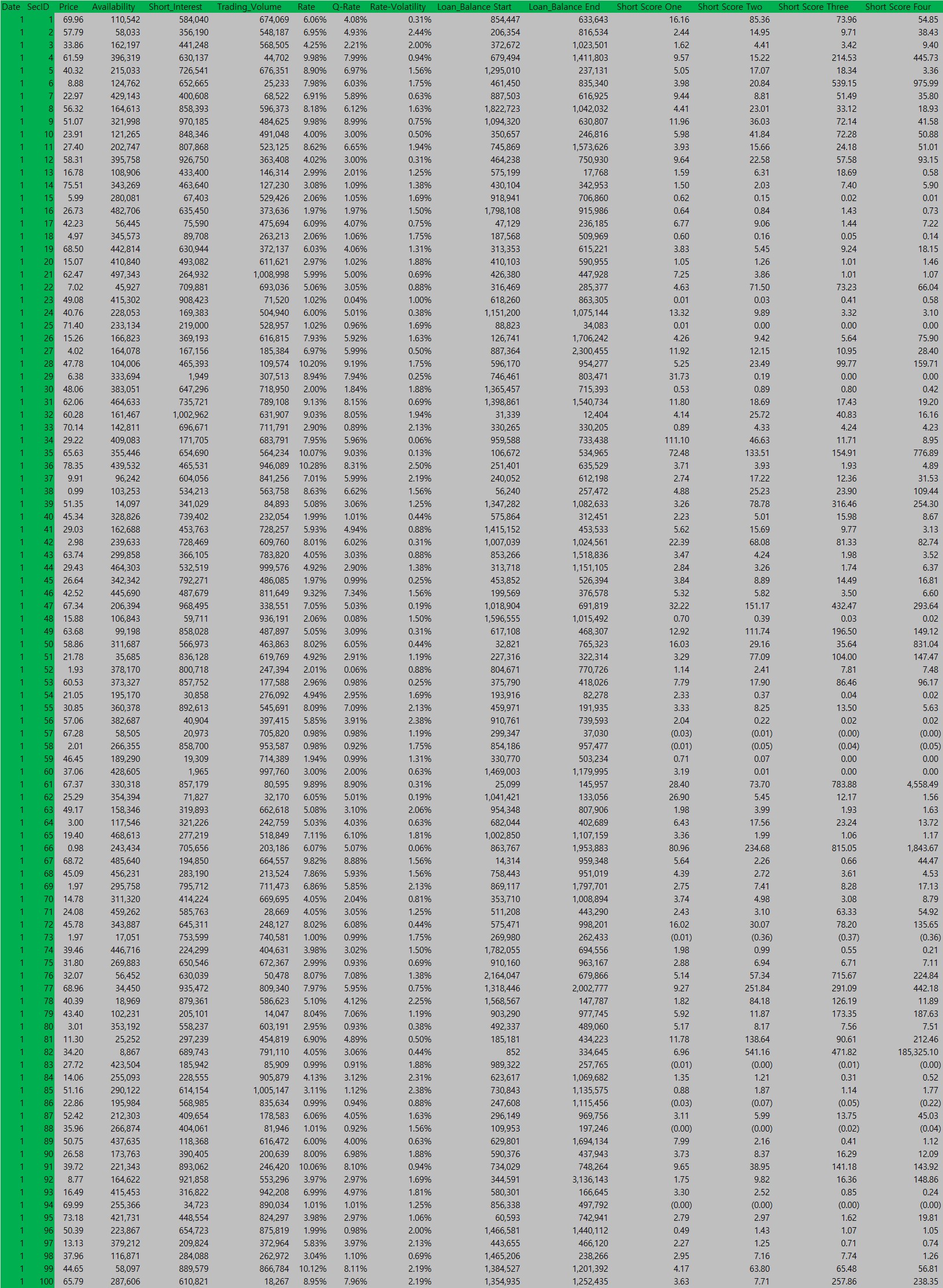}

\caption{\label{fig:Short-Scores--First}Short Scores - First Day of Sample}
\end{figure}

In Figure (\ref{fig:Short-Scores--First}) we provide short scores
for the securities based on the values on the first day of the time
series sample. The columns represent the following information respectively:
date, security ID, price, availability, short interest, trading volume,
loan rate, alternate loan rate, rate volatility, loan balance at the
start of the time series sample or on the first day, loan balance
at the end of the time series sample or on the last day, short score
one (Eq: \ref{eq:2}), short score two (Eq: \ref{eq:3}), short score
three (Eq: \ref{eq:4}) and short score four (Eq: \ref{eq:5}). Here,
the short scores do not use moving averages for availability, short
interest and trading volume but use values on the particular day indicated.

\begin{figure}[h]
\includegraphics[width=17cm]{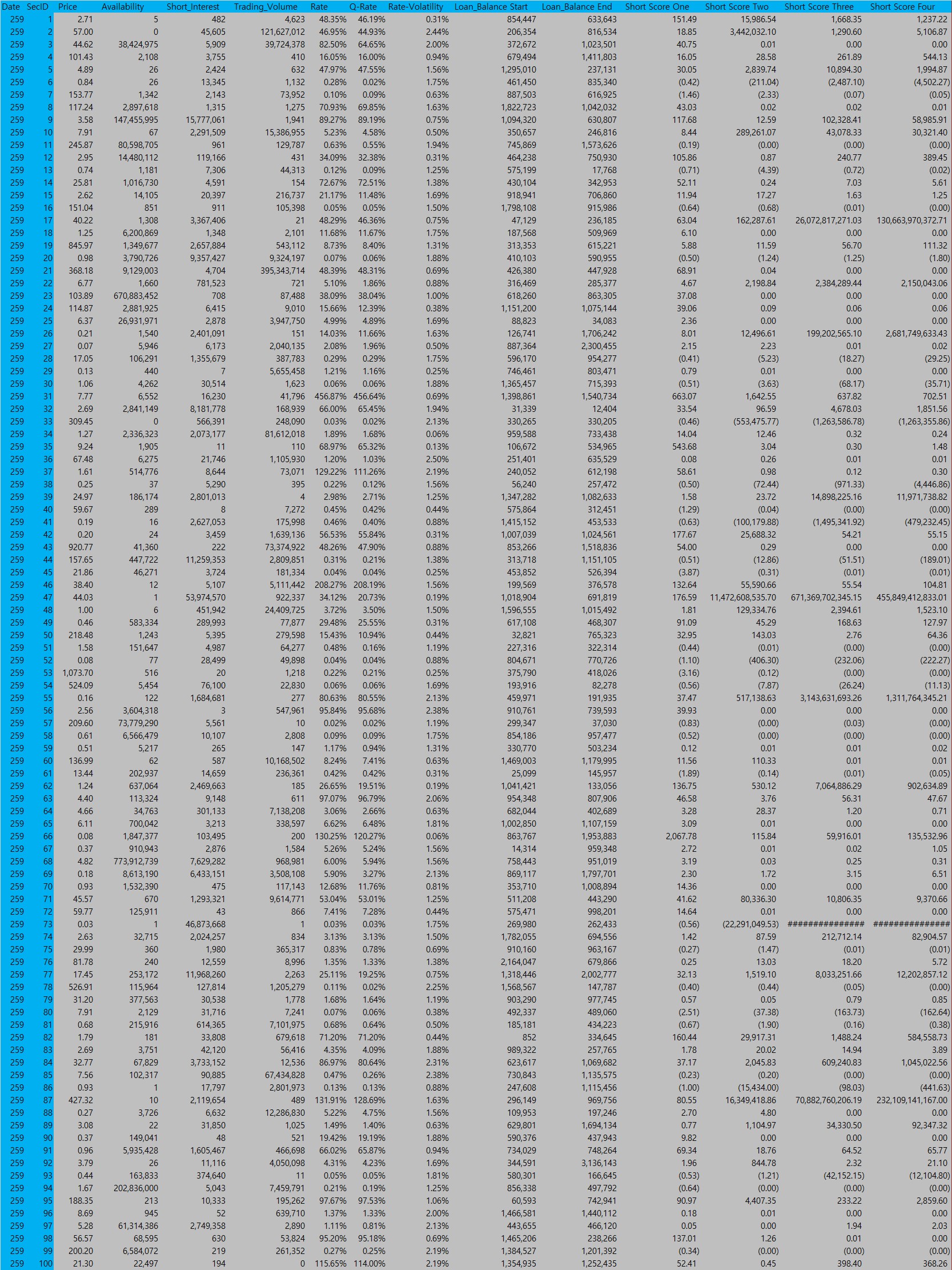}

\caption{\label{fig:Short-Scores--Last}Short Scores - Last Day of Sample}
\end{figure}

In Figure (\ref{fig:Short-Scores--Last}) we provide short scores
for the securities based on the values on the last day of the time
series sample. The columns represent the following information respectively:
date, security ID, price, availability, short interest, trading volume,
loan rate, alternate loan rate, rate volatility, loan balance at the
start of the time series sample or on the first day, loan balance
at the end of the time series sample or on the last day, short score
one (Eq: \ref{eq:2}), short score two (Eq: \ref{eq:3}), short score
three (Eq: \ref{eq:4}) and short score four (Eq: \ref{eq:5}). Here,
the short scores do not use moving averages for availability, short
interest and trading volume but use values on the particular day indicated.

\end{doublespace}
\begin{doublespace}

\subsection{Figures for Section (\ref{sec:Taming-the-Volatility}), Taming the
Volatility Skew}
\end{doublespace}

\begin{figure}[h]
\includegraphics[width=17cm]{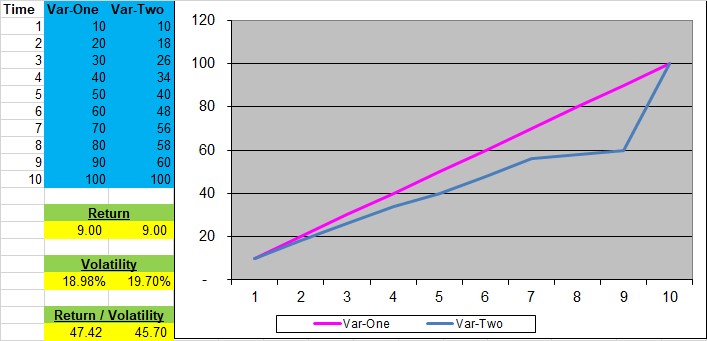}

\caption{\label{fig:Limitations-of-Volatility-1}Limitations of Volatility
- Upward Movement Penalized}
\end{figure}

\begin{doublespace}
In Figure (\ref{fig:Limitations-of-Volatility-1}), variable one and
variable two both have a steady upward path without a single down
movement. Variable two which has a higher volatility of 19.70\% ends
up having a lower return to volatility ratio and could get underrepresented
in a portfolio in comparison with variable one which has a volatility
of 18.98\%. Both variables have a return of 9\% over the time duration
under consideration.

\begin{figure}[h]
\includegraphics[width=17cm]{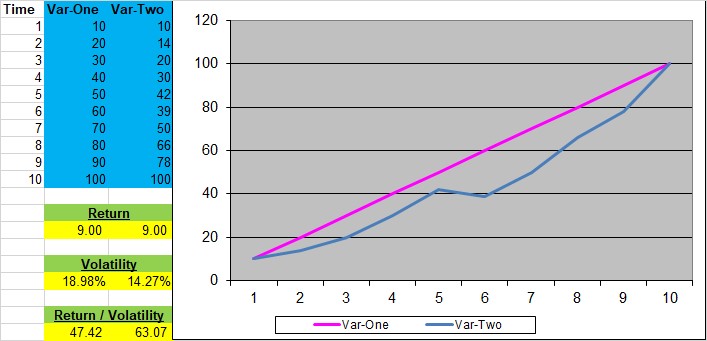}

\caption{\label{fig:Limitations-of-Volatility-2}Limitations of Volatility
- Downward Movement Not Penalized}
\end{figure}

In Figure (\ref{fig:Limitations-of-Volatility-2}), variable two has
a lower volatility of 14.27\% but has falling prices for some intervals.
In comparison, variable one does not have a single fall in price and
has a volatility of 18.98\%. Both variables have a return of 9\% over
the time duration under consideration.

\begin{figure}[h]
\includegraphics[width=17cm]{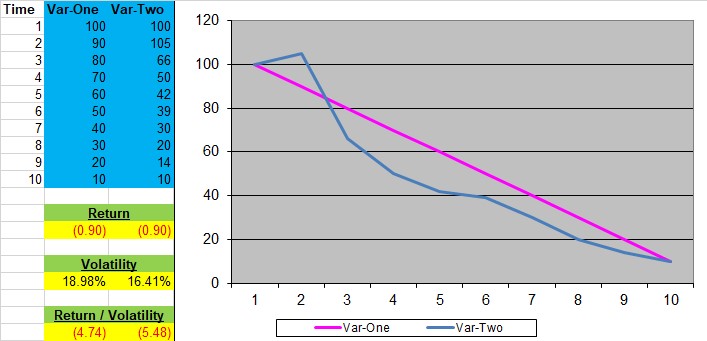}

\caption{\label{fig:Limitations-of-Volatility-3}Limitations of Volatility
- Downward Movement}
\end{figure}

In Figure (\ref{fig:Limitations-of-Volatility-3}), variable one is
falling more steadily than variable two. Variable one has greater
volatility of 18.98\% and hence can be seen as better than variable
two, which has a volatility of 16.41\%, from a return to volatility
perspective. Both variables have a negative return of -9\% over the
time duration under consideration.

\end{doublespace}

\end{document}